# Emerging Natural User Interfaces in Mobile Computing: A Bottoms-Up Survey

KIRILL A. SHATILOV, The Hong Kong University of Science and Technology, Hong Kong
DIMITRIS CHATZOPOULOS, The Hong Kong University of Science and Technology, Hong Kong
LIK-HANG LEE, The University of Oulu, Finland
PAN HUI, The Hong Kong University of Science and Technology, Hong Kong and University of Helsinki, Finland

**Abstract–** Mobile and wearable interfaces and interaction paradigms are highly constrained by the available screen real estate, and the computational and power resources. Although there exist many ways of displaying information to mobile users, inputting data to a mobile device is, usually, limited to a conventional touch-based interaction, that distracts users from their ongoing activities. Furthermore, emerging applications, like augmented, mixed and virtual reality (AR/MR/VR), require new types of input methods in order to interact with complex virtual worlds, challenging the traditional techniques of Human-Computer Interaction (HCI). Leveraging of Natural User Interfaces (NUIs), as a paradigm of using natural intuitive actions to interact with computing systems, is one of many ways to meet these challenges in mobile computing and its modern applications. Brain-Machine Interfaces that enable thought-only hands-free interaction, Myoelectric input methods that track body gestures and gaze-tracking input interfaces - are the examples of NUIs applicable to mobile and wearable interactions. The wide adoption of wearable devices and the penetration of mobile technologies, alongside with the growing market of AR/MR/VR, motivates the exploration and implementation of new interaction paradigms. The concurrent development of bio-signal acquisition techniques and accompanying ecosystems offers a useful toolbox to address open challenges.In this survey, we present state-of-the-art bio-signal acquisition methods, summarize and evaluate recent developments in the area of NUIs and outline potential application in mobile scenarios. The survey will provide a bottoms-up overview starting from *(i)* underlying biological aspects and signal acquisition techniques, *(ii)* portable NUI hardware solutions, *(iii)* NUI-enabled applications, as well as *(iv)* research challenges and open problems.

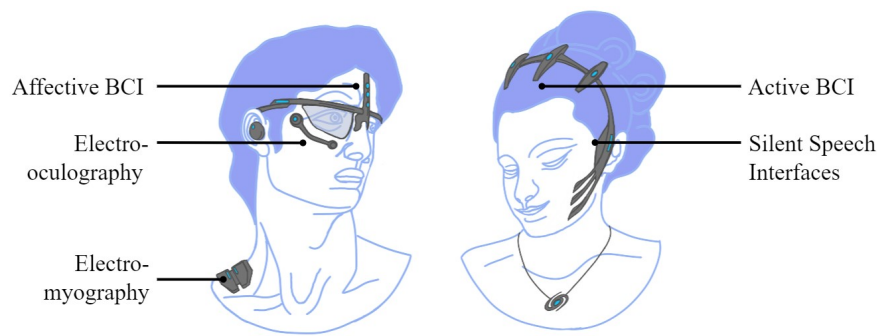

Conceptual illustrations of natural user interfaces for mobile users. Natural user interfaces transform collected signals from brain (BCI), muscles (Electromyography and Silent Speech) and eye movements (Electrooculography) to inputs on connected computer systems.

## 1 INTRODUCTION

Nowadays, there exist than 13 billion mobile devices in the world that are generating nearly half of internet traffic [60]. According to [74] there will be 17 billion mobile devices in 2023. In parallel, augmented, mixed and virtual reality (AR/MR/VR) technologies are showing intense growth: according to the authors of [105] the market size of global AR and VR technologies in 2019 is around 17 billion US dollars and it is expected to grow up to 160 billion USD in 2023. Mobile AR (MAR), as a technology in the intersection of AR/MR/VR and mobile computing, is also following the positive trend: the global revenue of the consumer mobile augmented reality application market is more than 3 billion USD in 2019, forecasted to reach 15 billion USD in 2022 [104]. This trend is backed up by the emergence and development of MAR frameworks from technological giants - Apple AR Kit [44], Google AR Core [70], Amazon Sumerian [140] - and newly introduced MAR capabilities of popular application-building frameworks - Unity [151], Unreal Engine 4 [66], and others. The wide adoption of MAR technology is held back by many factors. One of them is the cumbersome of controls [97]. The key limitation of existing input paradigms is the input bandwidth, which motivates the exploration of alternative approaches for Human Machine Interfaces (HMIs).

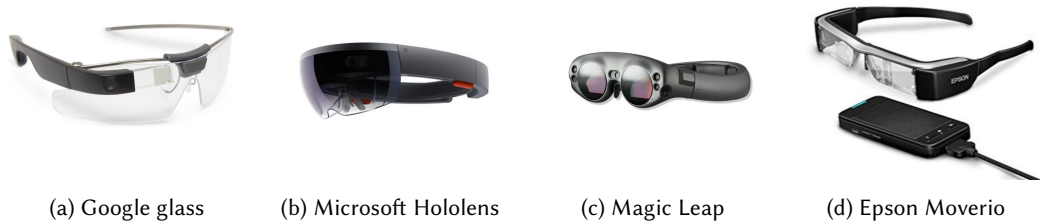

(a) Google glass  (b) Microsoft Hololens  (c) Magic Leap  (d) Epson Moverio

Fig. 1. MAR hardware: examples.

MAR can be divided into three categories: *(i)* Video See-Through (VST), *(ii)* Optical See-Through (OST) and *(iii)* Projective AR [57, 144]. VST captures the surroundings through a camera, and superposes the virtual world on top of the real one on a screen of a mobile device. OST devices display the augmenting elements directly in front of the user's eye on a semi-transparent screen [56, 58]. Several examples of OST head mounted displays can be found in Figure 1. Last, in Projective AR the virtual objects are projected onto a real environment. New HMI modalities suitable for the mobile usage are limited by several constraints [86]:

**Mobility Constraint.** The utilization of a modality should not interfere with other activities, e.g. walking in a crowd or doing sports.

**User-friendliness.** A modality should simplify interaction with mobile devices by providing convenient interfaces with high information throughput rates (ITRs).

**Non-intrusiveness.** A modality should not require the user's intent concentration on any input-related task; thus, users will be more aware of the surroundings.

**Non-invasiveness.** A modality should not rely on any invasive technique, e.g. brain implants that require surgery.

**Non-obstructiveness.** a modality should not introduce additional physical hardware that might constrain the user's physical mobility. In addition, hardware introduced by a modality should not obstruct the existing devices, for example MAR OST head mounted displays.

The two-way interaction between mobile users and mobile devices designated for MAR is not thoroughly designed. On the one hand, a number of micro-interaction approaches have been designed for displaying information to the users. For instance, wearable devices such as smartwatches can vibrate when the user is receiving a phone call or display the beginning of a message on their screen. On the other hand, the input approaches enabling mobile users to interact with the MAR devices are limited. The mobile devices supporting MAR, especially the wrist-worn and head-worn ones, have limited space for getting users' input [97]. Reacting to notifications and dismissing them, answering a call, or scrolling application feed relies on touch screen input, even though the required amount of input commands for these scenarios is limited to 3-4. The increasing adaptation of wearable with limited screen real estate pose a constraints to the users' input performances, where the users cannot manipulate digital objects directly [99].

For the case of desktops and laptops, users rely on bulky and tangible interfaces such as mice, touchpads and keyboards. In a theme of interaction design proposed by Klemmer *et al.,* desktop computers see the human users as an one-eyed, two-ear and one-handed creatures [89], which implies the users narrow down the two-way interactions to certain body parts and their capabilities. Information display primarily relies on the visual and auditory feedback, while the input on the graphical user interfaces rely almost exclusively on the users' hands. Therefore, it is necessary to explore alternative approaches for mobile devices considering that mobile devices serve as an attachment on users' bodies and, consequently, augment the human capabilities, one of the promising direction for input approaches is to exploit resources from the human bodies [95].

### 1.1 Natural User Interfaces

One way to tackle these challenges is to introduce specific physical controllers (e.g. Vive [61] and Oculus [150]), wearable keyboards [99], manipulation rings [97], gloves [98], and many others [51]. Another way is the development of Natural User Interfaces (NUIs) that function as natural extensions of person's cognitive and/or physical abilities [84]. NUIs circumvent the flaws of traditional input/output (I/O) interfaces, not only by helping disabled people to interact with computing devices, but also by augmenting healthy individuals, naturally extending their abilities with new input methods and modalities to interact with mobile user interfaces, virtual and real worlds in scenarios where traditional controllers fail or are highly constrained. Next, we introduce emerging NUIs that can accompany mobile devices.

**Speech Interfaces.** Speech-based input is a common example of natural user interfaces in the domain of human computer interaction (HCI). Speech as an input modality is widely used in mobile devices with the most popular application to be voice assistants (e.g., Siri [45] and Alexa [41]). However, spoken speech has several downsides as an input method: it can disclose sensitive information to both the surrounding listeners and remote speech processing servers. Wide adoption of speech-based interfaces is also inhibited by the fact that devices can overhear *any conversation* in the vicinity (eavesdropping issue) and that speech-based input systems listen *for everyone* in reachable area (impersonal devices issue) [86]. *Silent speech interfaces (SSIs)*, on the other hand, can provide silent, concealed and seamless ways of interacting with wearable devices and computer peripherals. SSI technology can be based on several modalities [69]: *(i)* video-based lip movement recognition, that has limited applicability in mobile computing; *(ii)* ultrasonic Doppler sensing; or *(iii)* decoding of facial muscles movements. Silent speech interfaces can be used for high-speed input of words and phrases within limited vocabulary using lightweight signal acquisition hardware [86].

**Gesture Inputs.** Gesture-based inputs is a second example of NUI deployed in contemporary computing systems. It varies from visual-based whole-body posture sensing, e.g. Microsoft Kinect [113], to detection of precise finger configuration [96], that can be achieved by using ultrasound produced and detected by off-the-shelf devices' speakers and microphones [102] or WiFi signals [160]. The abundance of various sensors in wearable devices brought other ways of gesture inputs. The authors of [167], for example, used a generic Photoplethysmography (PPG) sensor, which is used in most of the smartwatches and bands to measure heart rate, to identify gestures. However, most of gesture-based input systems are sensitive to interference and hard to deploy in mobile scenarios. *Surface Electromyography (sEMG)*,

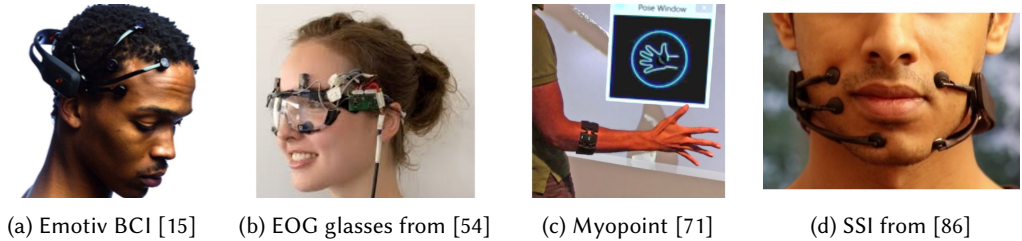

(a) Emotiv BCI [15]    (b) EOG glasses from [54]    (c) Myopoint [71]    (d) SSI from [86]

Fig. 2. Examples of existing NUIs

    a non-invasive method to quantitatively measure electrical activity in human muscles by estimating the electrical potential differences between muscle and ground electrodes, offers a robust and convenient method to recognize gestures as well. EMG measurements and analysis is used for medical, rehabilitation and sports purposes, alongside with HCI and prosthesis control by utilizing non-obstructive, lightweight sensors with the capability of wireless signal transmission.

  **Gaze Interaction.** *Gaze tracking* HMIs is a part of NUI paradigm as well. Visual-based eye-tracking systems [37] can precisely estimate where a user looks on the screen of a personal computer or head mounted display (HMD). Several solutions deploy small cameras on a frame of eyeglasses to track the eye sight, providing a tool suitable for mobile usage [34, 87]. Alternatively, *Electrooculography (EOG)* that measures corneo-retinal standing potential between front and back parts of an eye, can be used to track the eye position.

  **Brain Computer Interfaces.** Another modality is the Brain Computer Interfaces (BCIs, or Brain Machine Interfaces, BMIs). They allow sending a command to a computer using only the power of thought. Historically developed to provide a communication tool for locked-in or physically challenged patients, BCIs are evolving to a technology for a general consumer market [21, 25, 26, 32]. Modern non-invasive BCIs can provide a covert way of information input without any mechanical actions from a user [131].

Furthermore, different NUI modalities can be combined to build **hybrid** or **multi-modal** HMIs. Aiming to reduce the limitations of combined modalities, hybrid interfaces can achieve higher ITRs, better usability and universality [67, 94, 162]. NUI modalities can be divided into two groups: affective and direct control. When an **affective** modality is employed, user's mental state or physical condition is monitored. This information can be used to adopt a user interface [49], track the involvement into an activity [63], estimate emotional feedback [129, 145], and others. The second group, modalities of **direct control** are used to explicitly generate input signals, e.g. in the case of motor imagery control paradigm (Section 2.1.4), a user imagines the execution of certain motor activities. The applications of direct control modalities spans from spellers [137], to control of drones [155] and Internet of Things (IoT) devices [166]. Direct control modalities can be further subdivided into exogenous and endogenous depending on the source of stimuli. *Exogenous* HMIs use external stimuli to induce a certain activity. Steady State Evoked Visual Potentials (SSVEPs, Section 2.1.1), a BCI paradigm when a set of flickers is presented to a user, eliciting a response in the Visual cortex, is an example of exogenous HMI. *Endogenous* interfaces rely on an internal source of certain activity. Self-motivated gestures, speech or motor imagination can serve as examples of endogenous direct control NUIs.

  Natural User Interfaces, as a technology, face multiple problems: *(i)* the nature of biological signal limits effective bandwidth; *(ii)* signals are extremely susceptible to interference; *(iii)* acquisition hardware can be bulky and inaccurate; *(iv)* applications are limited. Moreover, the union of NUIs and mobile computing brings its own challenges: selection and combination of hardware, development and interoperability of software, competition for body space and energy supply.

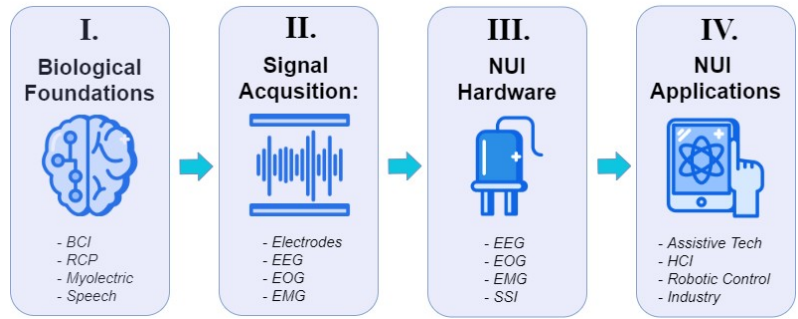

Fig. 3. Structure of the survey.

## 1.2 Contributions

This survey reviews emerging NUI modalities: Electroencephalography-based BCIs, EOG-based HCIs, and Myoelectric (including Silent Speech) HCIs, that are all non-invasive signal acquisition modalities (see Figure 2 for representative examples). All these techniques form the collective concept of E**x**G (E**E**G, E**M**G, E**O**G), and face similar challenges, such as the interference from external and internal noise, the challenge of energy and time efficient analysis of the recorded time series and many others [39]. The survey follows a bottoms-up approach to discuss the mentioned NUIs and their existing and potential applications. The aim of this survey is to bring together mobile computing and novel NUIs, such as BCI, myoelectric and gaze interfaces, serving as a bridge between areas. In general, biological foundations of Brain-computer interfaces are discussed, for example, in [133], paradigms design in [137]; myoelectric control interfaces in [127]; aspects of silent speech and its applications in [85, 114, 139]; hybrid interfaces in [75]. Other works have reviewed interaction techniques and their limitations for mobile systems: input methods in MAR [57]; limitations of interactions in smart glasses [97]; haptic technologies for mobile devices [51]. Several attempts were done in order to combine BCI and AR/VR [43, 129, 131, 144], BCI and drone control [124]. We outline the potential and existing NUIs paradigms from existing work that can be effectively introduced to use as input methods for mobile devices.

The rest of this manuscript is organised as follows: Section 2 the biological and physical background of the discussed modalities, explaining neurological phenomena of BCIs, biology behind EOG technology and physiological aspects of myoelectric currents in the body. Section 3 discusses mobile hardware for signal acquisition, providing examples of portable consumer and research grade devices. Section 4 provides an overview of existing NUI-enabled applications followed up by a presentation of challenges and open problems in Section 5. Section 6 concludes the survey outlining challenges of using NUI in mobile computing, as well as potential applications and emerging usage scenarios. Additionally, we provide three appendices with detailed information regarding portable EEG headsets (Appendix A), portable fNIRS systems (Appendix B) and wireless EMG sensors (Appendix C). The structure of the survey is presented in Figure 3.

| Band | | Range | Associated activities |
|------|---|-------|----------------------|
| *Delta* | 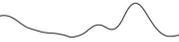 | 1 - 4 Hz | Deep sleep, relaxation, meditation |
| *Theta* | 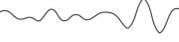 | 4 - 7 Hz | Light meditation, Rapid Eye Movement "dream-mode" sleep, information processing, learning, and memory recall |
| *Alpha* | 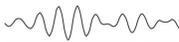 | 8 - 12 Hz | Deep relaxation, rest, daydreaming, meditation; sensory, motor and memory functions |
| *Beta* | 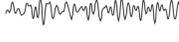 | 12 - 25 Hz | Consciousness, alertness, problem solving and decision making |
| *Gamma* | 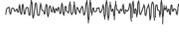 | > 25 Hz | Cognition, information processing, attention span and memory |

Table 1. Brain waves [133].

## 2 BIOLOGICAL FOUNDATIONS

In this section we briefly introduce biological phenomena and mechanisms upon which various NUI modalities are based: from the human brain and its properties that can be used in affective computing and to input multiple control commands; cornea-retinal potential that is measured by EOG; to the nature of myoelectric currents in muscles and speech production.

Additionally we introduce concepts of signal acquisition techniques and their metrics. We aim to outline NUI paradigms suitable for potential usage with mobile systems and unconstrained scenarios that can deliver acceptable signal quality according to the introduced metrics.

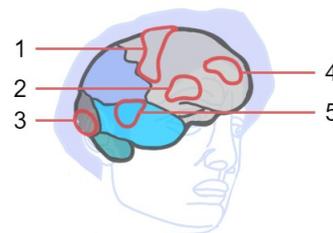

### 2.1 Brain-Computer Interfaces: Neurological Phenomena

The human brain contains 100 billion *neurons*; each neuron is an electrically excitable cell that communicates with others alike via specific connections, called *synapses*. Interconnected neurons form *synaptic*, or *neural* networks [73]. Once the sum of all incoming charges from synaptic connections reaches certain threshold, neuron passes the signal forward; this process of in-cell electrical potential rising and falling is called *action potential*. Rhythmic patterns of action potentials of neurons in the brain are referred to as *neural oscillations* or *brainwaves*.

Fig. 4. Brain lobes and areas: 1 - Motor cortex, 2 - Speech area, 3 - Visual cortex, 4 - Concentration and planning, 5 - Language.

Patterns of different frequencies are related to various types of activities, e.g. oscillations at frequencies of 8 - 12 Hz are called *alpha* waves and usually associated with the relaxed state of mind, meditation or daydreaming. Frequency ranges with related activities are presented in the Table 1. Affective state BCIs are based on analysing oscillation patterns by extracting dominant frequencies from the recorded brain activity. The brain can be subdivided into the following regions known as the *brain lobes*: Frontal, Pariental, Temporal, Occipital. The outer layer of the brain, which is called *cerebral cortex*, can be divided into smaller areas, each of them is associated with a certain activity or function [133]. Lobes and several BCI-related areas are depicted in Figure 4: MI-related potentials (subsection 2.1.1) are observed in the motor cortex; visually evoked (subsection 2.1.1) - in the visual cortex; attention related potentials such as P-300 (subsection 2.1.3) - in the frontal lobe.

The design aim of Brain Computer Interfaces is to map oscillating, endogenous or exogenous brain activity in specific areas to commands for a computing system. The speed of this conversion can be evaluated with *Information Transfer (or Throughput) rate (ITR)*. According to [141] ITR can be defined as follows:

$$ITR = \frac{B}{T}, \quad (1)$$

where $T$ is an average time required to input a command, and $B$ is a average amount of bits, which in turn can be defined as follows:

$$B = \log_2 N + P \log_2 P + (1 - P) \log_2 \frac{1 - P}{N - 1}, \quad (2)$$

where $N$ is a number of possible commands, $P$ is a selection accuracy ($1 - P$ represents the error frequency). Information transfer rates define how effective and quick users can utilize certain interfaces or communicate with each other. For example, the estimated ITR of human speech is roughly 40 to 60 bits/sec, or [2400; 3600] bits/min (bpm) [136].

*2.1.1* **Visually Evoked Potentials (VEPs)**. Visually Evoked Potentials provide the highest ITRs among contemporary BCIs [137]. BCIs based on VEPs belong to the group of *attention-based* BCIs - where the user has to focus on, or pay advert attention to one of the presented stimuli. Such stimuli can be a set of flickering markers or multiple haptic devices vibrating on different frequencies. Attention-based BCIs are sometimes also referenced as Selective Attention (SA) BCIs [94].

*Steady State Evoked Potentials (SSVEPs)* are the potentials evoked in the human brain by attending visual stimuli flickering at some rate. If multiple stimuli flashing at different rates are presented to the patient, the stimulus, that the patient attends to, will elicit a larger response amplitude than the stimuli that the user ignores. A typical SSVEPs interface with multiple stimuli is presented in Figure 5. Every flashing marker can represent buttons or any other UI elements. SSVEPs are distinguishable when analyzing the frequency spectrum of the brain of both adults and children [123].

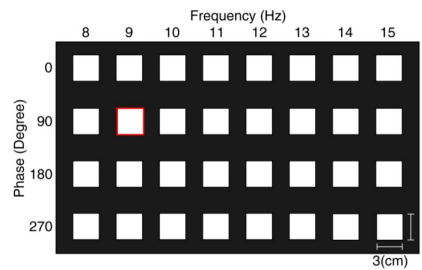

Fig. 5. SSVEPs input UI. A matrix of flickering markers from [118]

Usually, frequencies withing range [5; 50] Hz are used for controlling the flashing of markers. In [123] 6.2 Hz, 7.7 Hz, and 10 Hz are used; another study [101] used 9 flickers of frequencies from 27 Hz to 43 Hz with an incremental step of 2 Hz. The precise values of frequencies are chosen in order to illicit higher and more distinguishable from each other responses; additionally, the chosen frequencies are less likely to elicit photo-induced seizures [123]. Bigger number of presented markers, and thus higher ITR, can be achieved using not just various frequencies, but also different phases [118] at the same frequency. The time interval of a patient paying attention to a visual flicker required for accurate frequency detection is below 5 seconds [101, 118, 123]. The bigger the time window used for analysis, the better accuracy can be achieved by SSVEPs-based BCIs: from 55% within 0.75s to 76% within 2.25s in [101], from 85% within 1s to 93% within 4s in [118]. It is clear that longer time windows limit the ITR, making HCI system less interactive and responsive, thus longer attention spans are required.

SSVEP-based BCIs are widely adopted mostly due to the fact that they provide high number of available input commands (up to 32 in [118]) and require less concentration and mental efforts from users. On the other hand, constantly flashing, large and spacious visual stimuli might be considered tiring and irrelevant to the spelling task itself making some users feel fatigue and annoyance [107]. To address this issue other types of visually evoked potentials are considered for usage in non-invasive BMIs, such as *Miniature Asymmetric Visual Evoked Potentials (aVEPs)* [158]. Miniature aVEPs arise when very small and inconspicuous asymmetric lateral visual stimuli are presented in the peripheral vision, which is an area outside of 2 degrees of eccentricity of human visual field.

Presented peripheral stimulus subtends only 0.5 degrees of visual angle at an eccentricity of 2.1 degrees as depicted in Figure 6. Such a stimulus induces a different response in two hemispheres of the visual cortex, activating an area as small as 1.6 $mm^2$ [158]. It is also claimed that such a stimulus will be more compatible with AR/MR/VR scenarios as it allows user to explore surroundings, observe environment and use interface components without a need to focus on any particular task-irrelevant markers. However, it might be challenging to detect induced miniature potentials ($0.5\mu V$) using mobile EEG hardware.

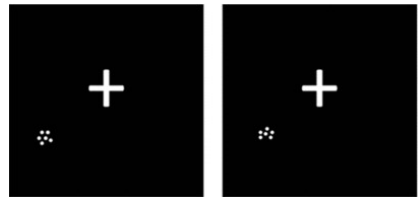

Fig. 6. aVEPs: the cross represents the gaze direction, points - stimulus itself (from [158])

*2.1.2* **Other Evoked Potentials**. A similar concept of paying attention to a certain stimulus is exploited in *Steady-State Somatosensory Evoked Potential (SSSEP)* [94] and *Auditory Steady-State Response (ASSR)* [122]. In case of ASSR several auditory stimuli each with different frequency or differently modulated amplitudes are played. The one that the patient pays attention to evokes bigger response in Auditory cortex. Typically, ASSR BCI are used in case when locked in patients can not control their eye movements; ITR of ASSR is below 5 bpm. Similarly, SSSEP paradigm estimates the brain's responses to a tactile stimulation. SSSEP can be used in hybrid BCIs as one of the complimenting modalities [94].

*2.1.3* **Event Related Potentials (ERP)**. Another example of attention-based BCIs is *P-300* paradigm. According to this paradigm a matrix of icons, which can be alphabetic symbols or any control UI elements, are presented to the patient, that focuses on one desired input. Each element of the matrix is intensified (or highlighted) in a random sequence. The elicited potential happens after 300 milliseconds (hence the name) after the element is intensified. P-300 potential is most clearly observed in frontal lobe.

*2.1.4* **Motor Imagery (MI)**. As an example of endogenous BCI, the MI paradigm relies on detecting asynchronous irregular patterns that emerge when humans willfully execute or imagine motor activity. Volitional imagination of moving a certain part of the body - tongue, left or right arm or feet - can be tracked by the increased cortical activity in the area of Motor Cortex designated to control the involved muscles. MI-based BCIs can be used to input one of several commands (usually binary left/right) at a time, thus they have limited ITRs [94]. Usage of

Fig. 7. P-300 input UI [137].

MI BCI is an intuitive approach to control drones or robotic manipulators: the desired direction of the movement can be perfectly harmonized with the direction of the imagined motor activity [124].

## 2.2 Brain-Computer Interfaces: Signal Acquisition modalities

A recording technique (or imaging modality) can be characterised by several factors:

**Spatial Resolution.** Characterizes the precision of measurements with respect to space and defines the ability of recording techniques to differentiate two objects. Spatial resolution of the acquisition techniques can be measured in *bits*, defining the range of values of the recorded signal.

**Temporal Resolution.** Characterizes the precision of measurements with respect to time and the ability of an imagining modality to distinguish two measurements in time. Temporal resolution is measured in $Hz$ and shows how many samples are read per second. There is often a trade-off between the temporal and spacial resolution [133]. Both resolutions define the required channel bandwidth to transfer data and the required storage.

**Signal-to-Noise Ratio (SNR).** Describes how useful is the information in the acquired signal and it is defined as:

$$SNR = \frac{P_{signal}}{P_{noise}}, \tag{3}$$

Where $P_{signal}$ is the power of meaningful information and $P_{noise}$ is a power of unwanted information in decibels.

Brain activity can be recorded using multiple techniques:

**Electrocorticography (ECoG).** Is an invasive imagining technique, where needle electrodes are inserted directly into cortex. It can capture a state of a single neuron, having the highest spatial resolution [133]. It has also a better SNR, than the other modalities, but the fact that electrodes have to be surgically implanted into skull makes the adoption of this technology questionable.

**Magnetoencephalography (MEG).** Is a non-invasive recording modality that measures magnetic fields induced by the electrical activity of synaptic networks. Even though MEG offers high spatio-temporal resolution, the required hardware is too expensive and bulky, making it impossible to deploy in mobile scenarios [133].

**Functional Magnetic Resonance Imaging (fMRI).** Detects the changes in magnetic field related to the blood flow. Although this modality provides high spatial resolution and is highly reliable, the offered temporal resolution is low ($1 - 2s$), delay is high ($3 - 6s$) and the hardware is stationary and requires a dedicated facility [133].

**Functional Near-infrared Spectroscopy (fNIRS).** Quantifies the concentration of specific chemicals in brain tissue using reflected near-infrared light. It measures, so called *hemodynamic response*, when blood flows towards certain active areas of the brain. As a technology fNIRS offers low cost of hardware and high portability, but it's spatial resolution is around $1 cm^2$, and hemodynamic response has a considerable lag [133].

*2.2.1 Electroencephalography (EEG).* EEG measures the electrical charges of the synchronized synaptic activity through electrodes placed on the scalp; a typical electrode consists of metal plate and amplifier. To describe the electrode placement over the head, international *10-20 system* is used. Electrodes are placed at a distances of 10 or 20 % (hence the name) of length between fixed reference points on the scalp (See Figure 8). The choice of electrode's location affects the BCI paradigms that can be used, for example MI-related potentials are best detected on top of the scalp in the Pariental area. EEG provides a high temporal resolution, i.e. it can detect changes in milliseconds, providing sampling frequencies up to 20 000 Hz (typical sampling rate spans from 250 to 2000 Hz). The fact, that hardware required for EEG measurements is affordable, has lead to its wide adoption and development. The potential of a single neuron is too small to measure using non-invasive techniques, thus a single electrode measures the sum of synaptic charges large enough to be propagated through different layers: brain itself, dura mater (three layers of the protective tissue around the brain), skull, skin, electrode gel (if applicable). The larger the pool of neurons involved in synchronous activity, the stronger the electric field they produce the clearer signal is acquired [82]. EEG is not devoid of limitations: it can only measure charges large enough to be propagated and measured; charges quantified on the scalp are smeared by the volume conduction effect. Moreover, charge propagation is affected by the skin conductivity which depends on the amount of dead cells, oil and water contents, sweat etc. Very important thing to consider in EEG systems is a noise factor: *external* coming from electrical devices around the recording hardware and *internal* noise - rib cage expansion, breathing, blinking, EOG and EMG artifacts, any other kinds of movements. Passive shielding of room and cables alongside with active electrodes (when amplifier is placed as close as possible to the recording surface) are used in order to minimize external noise. For reducing internal noise experiments are run in controlled environment and patient's behavior is strictly limited by experiment procedure. Additionally electrodes should *settle* first, allowing the potential of each electrode to reach a steady state before collecting data[82]. All these restrictions are making outdoor and mobile usage of EEG very limited: any movement of a user will create recording artifacts and every electronic device in the vicinity will affect the recorded electromagnetic field.

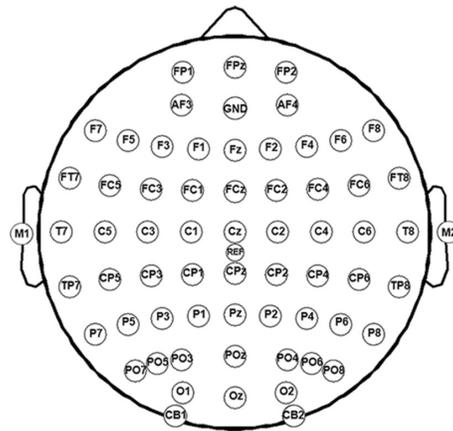

Fig. 8. Electrode locations according to the international 10-20 system (from [158]).

## 2.3 Electrooculography (EoG)

EoG is a non-invasive technique for estimating the permanent electrical potential difference between the *cornea* and *ocular fundus* of the human eye as shown in Figure 9 [91]. The eye can be represented as an electric dipole with a positive pole at the cornea and negative - on the retina. Change in the direction of gaze, controlled by eye muscles, transforms the configuration of the dipole, resulting in EOG potential. The detectable changes of the potential in range from 0.1 to 2 mV within 500 ms time interval, on the frequency of 0-30Hz and resolution of one degree can be detected by electrodes placed on forehead, temples and upper side of the cheek [47]. It is assumed that relation between the generated potential and eye movement is linear [157], however due to the fact that tissues around the eye are not distributed uniformly, the propagated signal measured on the surface can be considered only as an approximation of the biological reality [91]. In addition, EOG signal is influenced by the noise generated between the electrode and and the skin, myoelectrical activity of facial and eye muscles, and metabolic state of the tissue; potentials generated by the vertical eye movements are also sensitive to movements of the eye lids.

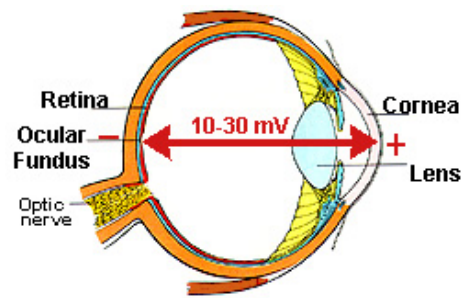

Fig. 9. Electrical potential measured by EOG (from [91])

## 2.4 Muscular Myoelectric Activity

Motor intentions, independently of their motivation (e.g. speech articulation, balancing or intended gestures), emerge in Motor area of the human brain. The control signal goes through submotor cortex, spinal cord and peripheral nervous system towards a motor neuron innervating a muscle. Nerve impulse that is being conducted through neuromuscular junction triggers the release of neurotransmitter into the synapse of motor neuron. The consecutive propagation of action potential causes the flow of ions, that generate a time-varying myoelectric potentials. As a result of the described process, target muscle contracts in a desired fashion[86]. Elicited myoelectric patterns can be observed using electromyography (EMG) invasively via electrodes incut directly into muscle, or non-invasively using surface electrodes. Surface Electromyography (sEMG) uses electrodes placed on top of specific muscle or electrodes organized as a grid [139]. Signals are measured as a difference between recording electrode(s) and ground (or reference) electrode, or utilizing several active electrodes, as in many commercial EMG devices [28, 36]. Similar to the EEG, recorded sEMG signal is a composition of potentials from several neurons, also smeared by the volume conduction through tissues, skin and skin-electrode interface.

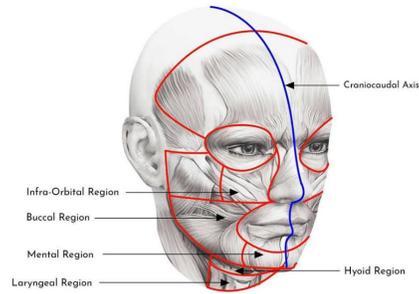

Fig. 10. Muscle regions involved into speech production (from [86])

*2.4.1 Myoelectrical Aspects of Speech.* Speech, as a process, can be considered as a transformation of *chemical* neuronal activity into *acoustic* signals via several stages: brain activity inside Broca's area triggers *electrochemical* impulses in peripheral nervous system that, in turn, controls jaw, lips, tongue, larynx and others in order to produce *mechanical* vibrations of a certain frequencies causing acoustic waves. It is interesting to note that myoelectrical signals in ortofacial muscles appear approximately 60 ms before the actual articulation [139]. Whispered, murmured and normal speech are processes that result in vocal vibrations while silent, imagined and inner speech do not produce acoustic outputs [139]:

- **Silent Speech.** Is a process when articulating muscles (see Fig. 10) are operated the same as in case of normal speech, but the respiratory activity is suppressed. Silent speech can be captured using sEMG, as well as using imagining techniques. To a certain extent, silent speech does not provide sufficient privacy as it exposes visual muscular articulation, so that an adversary can decode it.
- **Imagined Speech.** Occurs when both muscular and respiratory are suppressed. Imagined speech is similar to Motor Imagery BCI paradigm, and can be observed at neural level. Detection of neural activity related to imagined speech and decoding of obtained signal is a special case of BCI [114].
- **Inner Speech.** Is defined as internalized process when one thinks in pure meanings. It can be explained as verbal thinking, self-talk, inner voice or dialog. The assumption is that, inner speech is difficult to observable even at neural level, as it is related to cognition [139].

## 3 NUI HARDWARE

This section discusses the NUI hardware solutions offered by Industry and Academia. First, we introduce the mechanics of electrodes, their types and application cases. Then we provide a brief overview of portable EEG solutions, followed up by several examples of mobile fNIRS acquisition hardware (which is clearly the exception from ExG paradigm). After that, EMG and EOG hardware solutions are discussed. We also mention several hardware open source platform suitable for research or non-Industry deployment and usage. In this section we consider only wireless devices, that do not require wired connections for power or data transmission. Most of them can last for several hours autonomously, some (e.g. number 18 from Table 5) can be even charged from power-bank while recording. The target activity of this devices is to acquire biological signal in mobile scenario not restricting user's mobility.

We also show the API those devices expose to users, developers or researchers. We consider a number of channels ($N$), temporal resolution of each channel ($TS$, in Hz, equivalent of $sec^{-1}$), spatial resolution (SR in $bits$) and transmission protocol, e.g. Low Energy Bluetooth. Thus the effective bandwidth ($B_E$, in $bits/sec$) of utilized communication channel can be represented as follows:

$$B_E = N \cdot SR \cdot TR \tag{3}$$

The effective bandwidth influences the choice of transmission protocols, the requirement for channels to offload computations related to bio-signal processing. Classification software and algorithms that are intended to use in real-time should be capable of processing the dataframes formed by hardware, or employ downsampling techniques. It is worth noting that, regarding the sampling frequency of presented devices, according to the Sampling Theorem, when sampling the signal of frequency $f$, the sampling frequency should be twice as high, i.e. $2 \cdot f$. Thus, for example, devices whose sampling rate is 256Hz can reliably capture oscillations on frequencies ≤128 Hz.

### 3.1 Electrodes

As it was stated before, all introduced NUI modalities are based on recording techniques utilizing electrodes, that referenced as *ExG* [39] which is shorthand for EEG, EMG, ECG and EOG. Electroglottography (EGG), which is a non-invasive technique of monitoring vocal fold vibration by sensing the electrical conductance between two electrodes placed on the neck, can be also considered as one of representatives of ExG. There are multiple types of ExG electrodes, and they can be classified by the following criteria:

- **By Contact Type.** *Wet* electrodes require application of conductive gel or paste between the surface of the scalp and electrode to eliminate obstacles for charge propagation. Such electrodes can't be considered for everyday usage in consumer grade hardware: gel or paste turns greasy, deployment type takes up to one hour. Wet electrodes provide high signal quality and are used in medical grade hardware [133]. *Dry* electrodes require no conductive substance, electrode is contacting with a skin or hair directly. The air space is minimised by the pins on the end of electrodes. Some dry electrodes (e.g. OpenBCI, number 14 in the Table 4) contain springs pushing the pins towards the scalp. Dry electrodes might cause discomfort and irritation, continuous usage is questionable [133]. Another type of the electrodes is *semi-dry*: their Tips can slowly release a tiny amount of electrolyte liquid to the scalp. Emotiv Insight 5 (number 2 in the Table 3) is one of a few examples of hardware with semi-dry electrodes. The need to refill the syringe inside every electrode can be also considered as a disadvantage slowing down wider acceptance by users.
- **By Applicable Surfaces.** To provide a better signal in various conditions different types of electrodes are used: for acquisition of the signal through *hair* a set of pins on the end of electrodes is used; *skin* electrodes usually have plain surface, while *universal* electrodes can provide decent quality of records from any surface.

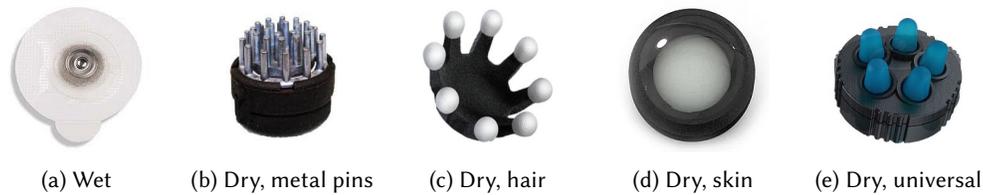

| (a) Wet | (b) Dry, metal pins | (c) Dry, hair | (d) Dry, skin | (e) Dry, universal |

Fig. 11. Different types of electrodes: examples.

**By Amplification Capabilities.** *Active* electrodes contain an amplifier with factor of 1-10 to minimize the interference, while *passive* electrodes do not contain such an amplifier on-site.

**By Material.** Usually electrodes are made from gold (*Au*) or silver (*Ag/AgC*). Gold electrodes are suitable for measuring different modalities (e.g. EMG, ECG), while silver ones proved to be more reliable to record EEG signal of low (below 1 Hz) frequencies [133].

Examples of electrodes of different types are depicted on the Picture 11.

## 3.2 Brain-Computer Interfaces

Tables 3, 4 and 5[*], in appendices A, B and C respectivelly, present the up-to-date set of portable EEG headsets. Several criteria might be introduced to define portability and user-friendliness, we adopt metrics from [55]:

**Ease of Deployment.** Preferably, dry or (semi)dry electrodes are should be used, providing the deployment within reasonable amount of time (< 5 min). Ease of deployment also implies the speed and required qualification efforts to obtain, record and analyse a signal from a recording device.

**Portability.** The EEG hardware should be capable of producing signal good enough for better than random guess analysis in non-shielded mobile scenarios in naturalistic settings with a lot of physical and electromagnetic noise.

**API Availability / General Applicability.** Classical medical EEG headset system come with proprietary highly-integrated hardware and software environment, that provide fixed list of certain analysis methods. A device should expose API with access to raw data. Regarding this metrics, some systems, such as EMOTIV (number 2 in Table 3) requires paid subscription to access raw EEG data and still limits the amount of recordings per day.

In addition to EEG acquisition devices presented in Tables 3, 4 and 5 there exist several projects for developing EEG acquisition hardware supported by community. For example, ModularEEG [126] is a community of enthusiasts that aim to build low cost EEG acquisition hardware and build software ecosystem around it assisting signal collection and classification. Hardware device, based on this platform, EEG-SMT [117] has been used in study [43]. Another example of a hardware platform for bio-signal acquisition is BITalino [52]. BITalino offers several sets of hardware for collecting EEG, EMG, ECG signals, as well as amplifiers and connectivity (Bluetooth or BLE) boards.

---

[*]In presented tables we used several simplifications: → is used to depict digital downsampling performed on board. For example, 1024 → 128 implies that signal is recorded at 1024 Hz, but the digital data stream contains 128 samples per second. **OR** is used to describe the variative capabilities of a single device: $500Hz$ **OR** $1000Hz$ means that a device can capture signal on both frequncies, 500Hz and 1000Hz. / is used to combine a several products withiong a certain product line with similar characteristics. For example, Cognionics QUICK (number 10 in Table 4) product line consists of three similar products, differing only in electrodes count.

Several studies accessed and compared the capabilities of consumer grade EEG devices. Authors of [135] compared the performance in terms of power spectra similarity of B-Alert X24 (number 17 in Table 5) and Enobio 20 (precise medical grade EEG headset) with consumer oriented Muse and NeuroSky Mindwave (number 6 and number 5 in Table 3 respectively). This study showed that Neurosky EEG obtains signal from electrode location Fp1 of similar power spectrum with medical grade hardware, while increased values of power spectra were detected in signal collected by Muse headset. Muse demonstrated highest relative variation across multiple acquisitions; both consumer headsets suffered greatly from artifacts such as eye blinks. Authors conclude that although EEG signal can be successfully collected by all devices, medical grade hardware offers better quality and more reliable signal suitable for many applications.

Another study [132] compared EPOC (number 1 in Table 3), Mindo Trilobite / Jellyfish (number 20 in Table 5), BR8PLUS (number 19 in Table 5) and g.SAHARA / g.LADYbird (electrode layouts for g.Nautilus, number 23 in Table 5) employing 24 participants performing multiple BCI-related tasks for 6 consecutive days. Devices were compared based on multiple criterias: SNR, proportion of recorded artifacts and frequency domain quality. The study concludes that mobile systems with wet electrodes are producing more stable signals of better quality than the systems with dry electrodes. Nevertheless, the quality of recorded signal using dry electrodes EEG was characterised by authors as "comparable and promising". Signal quality and research applicability of Emotiv Epoc+ (number 1 in Table 3) is accessed in [55]. Authors outline that despite several challenges, consumer-grade EEG headsets represent a useful addition to BCI research. In [100] overview of research projects in the area of BCI hardware is presented; authors conclude that existing BCI hardware is held back by many factors and they expect more emerging projects in this area tackling today's challenges.

### 3.3 Functional Near-Infrared Spectroscopy (fNIRS)

Although fNIRS aqucistion modality is clearly outside of ExG paradigm, in recent years portable fNIRS solutions evolved significatly. In Table 6 we present a short reference of up-to-date mobile fNIRS hardware for comparison with previously presented EEG helmets and devices.

### 3.4 Surface Electromyography

The mapping of the collected myoelectric signals to the user's gesture is complex and highly dependent on the quality of the collected signals and the selected classification method. sEMG devices are able to collect and amplify the signals generated by the human muscles and, depending on their processing and networking capabilities, process them and transmit them to other devices. There exist a wide variety of sEMG recording hardware aimed for medical purposes: solutions from BTS Bioengineering [53], MotionLabs [116], Otto myoblocks [1]. Overview of commercial consumer-grade EMG devices is presented in Table 7. Additionally there exist several hardware platforms suitable for sEMG signal collection, for example, PicoEMG [146], sensors of which can be organized as a mesh over muscles of interest, each sensor is capable of sampling data at 2000 Hz for up to 12 hours, weighting only 7 grams. Open source projects, such as MyoWare [77] offer only EMG electrodes with bio-signal amplifiers, that are compatible with Arduino [46] platform. It is expected that enthusiasts and researchers will build their own EMG sensing systems taking care of power supply and data transmission.

### 3.5 Electrooculography

As it was stated before, EOG potentials can be recorded by electrodes placed on both temples, upper cheek, and forehead. It also claimed that retino-ocular potentials are best detected on the lids and external canthi (the bone on the side of the eye) [91], however such a placement might be considered obstructive and irritative, limiting the range of eye movements. In some studies EEG hardware with electrodes placed on specified positions around eye is used to measure EOG potentials: g.USBamp (number 23, Table 5) in [107]; OpenBCI (number 14, Table 4) in [148]; or medical-grade Mobi8 by TMSI [152] in [145]. Many researchers have designed and built customised EOG-signal acquisition solutions: [47] employed wet electrodes and custom made amplifier; hardware described in [108] consists of two electrodes and Arduino-based amplifier; authors of [157] used 5 wet electrodes, in-house designed amplification hardware with Bluetooth module.

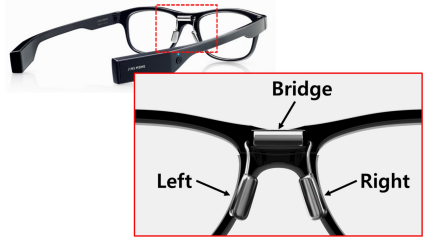

The authors of [54] present EOG glasses with 5 dry electrodes capable of detecting 8 different eye gestures. Built-in battery can maintain up to 7 hours of functioning. Similar glasses presented in [59] employing 3 dry Ag-AgCl electrodes, custom-made amplifier, signal conditioning unit and wired connection to a PC for signal processing. There are also commercial EOG products on the market, such as Jins MEME ES [111], that uses 3 dry electrodes (as depicted in Fig. 12) and employs additional 6-axis IMU. These glasses are reviewed in [81], authors present the methods of distinguishing several everyday activities based on sensors data, such as typing, eating, reading and talking. Another example of commercial EOG glasses is Imec Eye-tracking glasses [80]. It is mentioned that EOG sampling rate of 256 Hz is superior to any video-based eye-tracking system, as well as energy and cost efficiency of such a solution surpasses video-based systems. These glasses are equipped with battery allowing them to last up to 10 hours and a Bluetooth module for wireless communication.

Fig. 12. Electrodes placement of commercial EOG acquisition device

### 3.6 Silent Speech Interfaces

In order to obtain myoelectric signal related to speech production, sEMG electrodes are placed in strategic locations on the face and neck: [154] employed 5 sEMG electrodes sampling at 600 Hz; [62] used 8 wireless sEMG sensors that are a customized version of the Trigno wireless sensors [79] sampling at at 20kHz; in multi-modal approach to SSI, as described in [69], beside using Microsoft Kinect Depth camera, custom-built ultrasonic sensing device, authors used 5 pairs of sEMG electrodes from Plux [138] capable of sampling the signal at 500 Hz.

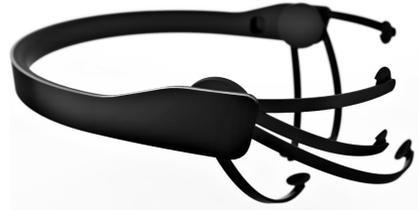

Fig. 13. SSI mobile wearable device [86]

There are number of design aspects that apply to a mobile SSI acquisition device in order to obtain robust sEMG signal for silent speech processing. First, electrodes should not change their position during various activities, e.g. walking or running, to maintain consistency of recorded signal *within one session*. Another consideration is that electrodes should be deployed at the same position every time device is put on to maintain the consistency *across multiple sessions*. Additionally positions of electrodes should be adjustable to be placed on top of the same muscles independently of facial shapes of *multiple users*. A device meeting these requirements [86] is shown in Figure 13: a wearable fixed around top part of the neck with electrodes placed on the ends of adjustable manipulators that in the same time provide desirable rigidity for steady signal acquisition. The described device uses 7 wet golden plated silver or passive dry Ag/AgCl electrodes; sEMG signals are sampled on 250 Hz. A fusion of silent speech device with self-contained breathing apparatus is presented in [85]. Such a system employs two wet EMG electrodes placed under the chin and on the left side of the neck. This work shows the compatibility of SSI with specialized equipment in cases where silent speech communication is most applicable and useful - noisy and harsh environments.

## 3.7 NUI Hardware Challenges

Although several studies evaluate the capabilities of modern consumer grade signal acquisition hardware [55, 132], there exist multiple challenges in the filed of bio-signal imagining modalities:

**Temporal Limitations.** Even though sampling frequencies of several EEG and EMG devices are high, instant voltage drops (for example those related to cognition) might not be detected by the acquisition hardware.

**Spatial Limitations.** The density of the electrodes is limited by obvious physical constraint on the head, arm or face. However for some applications, like precise gesture recognition, it is better to have dense electrode arrays over certain muscles[115].

**Signal-to-Noise Ratio.** Any mobile device is susceptible to internal and external noise. Leveraging the effect of electromagnetic interference and movement artifacts both programmatically and on the hardware level is important milestone on the way of delivering reliable and robust mobile imagining modalities.

**Energy Efficiency.** Management of a power supply for acquisition hardware and bio-signal processing facilities is crucial task in deploying user-friendly NUIs. The capacity of batteries that user can carry is physically limited, so all wearable devices will compete for power supply. Energy efficient offloading of computationally heavy tasks may reduce the need for power [142, 164].

**Signal Transmission.** Wearable devices and sensors are additionally constrained by networking interfaces. Delays, interference, collisions and protocol bottlenecks are challenges to address. [163].

**Users Comfort.** In the case of user comfort several problems arise: the whole system can be bulky or obtrusive or connected electrodes might cause discomfort. Dry electrodes are causing pain and irritation which makes them not suitable for long usage; while conductive gel used in wet electrodes leave sticky stains and requires reapplication in certain cases.

## 3.8 Hardware Challenges of Combining Mobile Output and NUI-related Hardware

As mentioned before, there are multiple issues to address when developing NUI-enabled for mobile computing. It is clear that BCI, EOG or SSI electrodes will compete for the space on user's head and/or face with AR glassed, HMDs and other output devices. Usage of head-mounted MAR systems, like Hololens, can block placement of electrodes in frontal and temporal lobes, while electrodes in central areas of the scalp can be placed freely. In some cases, electrodes can be placed around display of MAR or VR system's chassis, but it will most likely result in discomfort or recording artifacts and signal interference [131]. Systems like DSI-VR300 (Number 15 in the Table 4) and Looxid VR (Number 16 in the Table 4) are examples of how this issue can be addressed. Yet, this tight integration comes with its own shortcomings: the amount and configuration of electrodes is fixed (limiting the potential BCI scenarios); NUI hardware, as well as MAR or VR headset, can not be used separately; vendors can potentially limit the scope of compatible software, SDKs, AR/VR platforms etc. Another challenge is ergonomics in general. Combination of complex hardware can be bulky and require longer deployment time, MAR headsets, for example, and NUI-hardware can strain user's neck and head negatively affecting user's experience and shortening the usage sessions. Requirements for total weight and comfort might restrict the allowed weight and, as a consequence, capacities of batteries that can be placed on the user's head, thus creating another competition for the power supply.

## 4 NUI-ENABLED APPLICATIONS

In this section, we discuss original applications of BCI and EMG that assist individuals with special needs, such as people of motor impairment. Next, we present E*x*G applications and one recently proposed EMG-based silent speech wearable device. Finally, we introduce different use-case scenarios in various aspects such as biometrics and security, robotics, IoT devices as well as entertainment and gaming.

***BCI as an Assistive Technology***. Interpreted signals from NUI modalities can be used as an input to computing systems. BCI spellers are tools that enable the input of alphabetical characters for patients who lost their motor and speech functions. One of the most known examples of effective NUI applications is the Stephen Hawking communication tool, a speller that is used to assist people with special needs [114]. Words presented on the screen mounted to the wheelchair were selected using switch actuated by hand, head or eye motion. This system achieves an input rate of 15 words per minute. State-of-the art high-speed speller [118] based on SSVEPs reach 50 bpm by presenting to user 32 flickers of 8 different frequencies and 4 phases. To demonstrate the viability of miniature aVEPs authors of [158] built 32-alphanumerical speller, achieving ITR over 30 bpms. Deep learning solution utilizing one dimensional CNN used in [121] in combination with an EMG channel from self-designed EEG headset showed high accuracy of 99.2% with ITR comparable to the state-of-the-art (50 bpm). A review of up-to-date BCI spellers is presented in [137].

***EMG as an Assistive Technology***. EMG signals are mostly used for naturally controlling hand prosthesis, thus EMG-based NUIs are challenged to recognize more gestures simultaneously within less time and more accurately. The authors of [115] pushed the amount of classified labels to 21 by utilizing a compact mobile and high density EMG grid. The authors of [161] successfully identified ten gestures utilising CNN with a single convolutional layer. A significant contribution of that study is the consideration of how EMG signal changes over time, and how classifiers can be adjusted to the temporal variation of biologically originated signals. Besides being applied for gesture recognition directly, sEMG employed in other scenarios: The authors of [50], for example, aim to identify the exact finger being used for interacting with touch device (or any surface) and to measure the force applied, thus providing extra contextual information on HCI interaction. Among applications of this solution, the authors of [50] mention the possibility of turning any surface into a touch-enabled input device, advanced text marking and few others. The authors of [71] proposed a system of controlling any personal computing device utilizing the IMU embedded into Myo Band to control a cursor and gesture to input different commands, providing an alternative to mice.

***EOG Applications***. The goal of gaze tracking systems is to provide an intuitive general purpose HMI for users, including those with disabilities, such as locked-in syndrome and Amyotrophic lateral sclerosis (ALS). Such a system is build in [148]: recognising only two classes of activity (closed eyes and turn right and no activity) with high accuracy of 94% and ITR of 22 bpm was achieved. In the study of [157] eight directional eye gestures were recognised with 88.59% accuracy, achieving 32.42 bpm ITR.

In order to show the feasibility of the developed EOG acquisition hardware and eye gesture input, the authors of [54] designed a computer game, where users were repeatedly performing a set of predefined gestures. During the experiment there was no restriction on movement of the head and upper body. After the experiment around one third of all the participants reported that they had difficulties concentrating on eye gesture input task. A similar gamification technique is applied in [47]: authors built an LED panel and participants were asked to put the target to the destination position on the panel. Using fine-tuned thresholding, the authors achieved 80% accuracy for 5 available input commands over 500 ms time windows (thus ITR is 57bpm). As a potential application of the developed system authors mention the virtual keyboard or the wheel chair control. In order to extend the number of available input gestures it is proposed to combine several simple ones, e.g. blink and consecutive look right, so that more complex input protocol will improve transfer rates. The study of [63] proposed a comprehensive input protocol, combining data from IMU sensors embedded into off-the-shelf EOG glasses with EOG data directly. An ITR of 300 bpm with 500 ms per input was achieved. EOG can be used as an affective state interface: several studies [108, 168] employed EOG modalities to detect drowsiness; in [145] several emotions - happiness, sadness, anger, fear and satisfaction - were recognised purely based on the eye motion.

***Applications of Hybrid Interfaces***. Hybrid, or multi-modal, Human-Machine Interfaces, in the foundation of their design, aim to extend the number of available commands, increase the classification accuracy and reduce the decoding time [75]. Such HMIs can utilize advantages of certain modalities when combining them. One of the sources of inner noise in BCIs is electrical activity and mechanical artifacts caused by eye movements. Tracking of such an activity in some cases doesn't require additional hardware [107], so tracking data can be used as to filter the BCI signals, beyond just providing additional input modality [67]. The hybrid HMI presented in [107] combines EOG and EEG and is used to control humanoid and swarm robots. Presented solution uses one hardware device (g.USBamp) to record both signals simultaneously. The authors of [162] combined EOG, EEG, EMG to control a robotic soft hand. 21 palm gestures are recognised by analyzing EMG data with portable lightweight self-made device. Authors of Gumpy [149] shown the feasibility of the proposed hybrid toolbox by several applications: control of robotic hand using SSVEPs, prosthetic hand control based on EMG and 2-label motor imagery for general BCI. More importantly, they demonstrated the feasibility of multi-modal BCI based on the proposed toolbox by issuing commands for a robotic arm simultaneously by imagining movements (MI) and performing gestures (EMG). Undoubtedly, assistive NUI-based means of communication are improving quality of life of disabled people providing tools to interact with family, medical personnel and society.

***Applications of Silent Speech Interfaces***. EMG-based SSI has several advantages over other SSI modalities, that makes them more applicable: deployment of dry electrodes over the ortofacial muscles is fast and easy; less constraining hardware will cause less discomfort for a user; less training and adoption means fast enrollment; properly designed hardware will minimize errors and reduce environmental noises [86]. Several SSI applications can be outlined [139]:

> **Voice Prostheses.** Are designed to restore spoken communication for those who are challenged to produce audible speech. For example, in [110] sEMG SSI device is proposed in order to assist verbal communication for people with removed larynx: vocabulary of 2500 words is recognized allowing patients to have basic conversations.
> **Speech Therapy and Language Learning.** SSI can help capturing speech production abnormalities and errors; the provided multi-modal feedback can help patients learn and correct pronunciation [139].
> **Robust Voice Communication in Noisy Environments.** Developed for both, military and civilian contexts. Voice, distorted by external noises, masks or garments, can be reliably captured by SSI. An example of SSI assisting firefighting process is presented in [85], where EMG electrodes are embedded into breathing apparatuses.
> **Mute Spoken Communication.** Also called human to human silent communication, or synthetic communication. Once vocabulary of SSI is large enough and means of output are provided (e.g. bone conducting headphones), networks of people can exchange information within each other. Such systems can be deployed in environments where noises are prohibited or socially unacceptable [62].

**General HCI.** Study [86] outlines several potential HCI use cases: *closed-loop interfaces* when a user inputs non verbal commands into a system that provides feedback, for example, assisting with arithmetic calculations; *open-loop interfaces* when a user employs SSI to control smart devices, media or services, e.g. ordering a taxi, adjusting temperature and so on.

**Biometrics and Security.** NUI based biometrics has been used for many decades for now: face and fingerprints recognition systems are clearly the part of NUI paradigm. Believed to be secure, facial or fingerprint biometrics still can be forfeited or stolen using fingerprint casts and facial masks. Biometrics based on BCI, on the other hand, can not be replicated or applied involuntary. Although information transfer rates of most NUIs are relatively low, and primary application area of BCI is assistance of disabled, BCI based biometrics is emerging and promising application for society in general [93].

Password input in AR/MR/VR can be very reluctant and slow. EEG based on consumer grade devices can offer promising and secure tool for biometric authentication. A system that captures and classifies oscillating waves of the brain can be used by anyone and might be more convenient for people with disabilities, as it does not require subjects to move. Prior work establishes EEG signals recorded over certain time as feasible biometrics with high accuracy [93]. On the downside, EEG signal is a subject to change over time, even between 24-hours span, thus classifiers employed in EEG biometric systems should be robust and able to track the continuous change in a signal.

**Industrial and Commercial Applications.** MAR found its way into industry: it can be useful in diagnostics, maintenance, inspections, repair, training on-the-job and product design [57]. Most of this tasks usually require work with an instrument, thus it is inconvenient for a user to use touchscreen or any other conventional input device. As in other similar cases, NUI can provide hands-free input. In [43] authors proposed a mobile system assisting workers utilizing Epson Moverio BT-200 AR platform and SSVEPs paradigm. All software runs autonomously under Android OS in real time. BCI can be used to analyse users behaviour as it is able to track emotions and decision processes. In a study [128], users' emotional perception and emotions are captured and subsequently associated with different e-commerce activities at various phases of buying process within the e-commerce platform. Insights gained from such an analysis, e.g. positive or negative emotional reactions to a certain activities, can be further used to improve the design of the platform. Similarly, advertisement [159] and filming [134] industries can deploy BCI to detect users' attention and emotions, in order to understand the customer satisfaction adding a new dimension to customer studies.

**Control of Robots and Drones.** Study [155] combined SSVEPS and VR in order to control a quadcopter in virtual environment. Four SSVEPs stimuli were presented on the screen of HTC Vive VR helmet to control a drone in 3D virtual scene that was created using Unreal Engine 4. In [143], authors conducted several studies: they tested general feasibility of combining AR and BCI; explored different positions of SSVEPS input markers regarding the controlled object in AR; and applied their solution to control a robot. Studies [90, 124] present detailed overview of BCI applied to control of unmanned devices. Other modalities such as EMG and EOG can be also used to control drones and wheelchairs [88, 127].

**Control of Environment.** NUI can augment and expand traditional ways of interacting with real and virtual world in context of controlling smart home appliances. For example, BCI commands for controlling IoT devices can be input hands-free, silently, without disturbing other inhabitants of the environment, or when other modalities are not available [42, 166]. Also, the smart garments can be controlled by BCI [112]. In that case, BCI serves as sensory devices to detect the users' mental state and accordingly adjust the appearance of the smart garment. In this way, BCI-driven sensory can inform the surrounding people about your perception of surrounding and social environments, as well as display the emotions through color variations. Such an approach might provide a novel experience in the everyday social interactions.

***Entertainment***. The authors of [129] proposed a neuro-feedback driven AR application for kinetic meditation. In this study the intensity of alpha band is analysed to estimate user's mental state. The monitoring of mental state gives significant insights to game designers about the users' emotions [76] as well as the user experiences [64] throughout the gaming process. This state is then reflected by AR scene and surroundings are being adjusted accordingly in virtual reality. The study of [129] shows that participants described the combined AR/BCI experience as "interesting" and "promising" mostly giving the positive feedback. Neuro-feedback can be one of the tools to increase user's engagement, such as the gamification and augmented reality applications [130].

***Others***. The authors of [106] proposed a unique application for BCI. In their work, they focused on a human subject whose brain generated labels were used to train an image recognition network. Images at the rate of 4 Hz were shown to a patient wearing DSI-24 (Number 21, Table 5) and recorded EEG signal is then used to obtain the so-called *soft labels* that represent the confidence of a certain class. Finally, these labels along with initial pictures were fed into an image recognition network for training. For 2-labels classification this work outperforms the network trained on the manually labelled pictures (AUC 0.91 vs 0.89) within much less time required for labelling (22.8 minutes vs 257.8 minutes). Furthermore, the study of [49] focuses on utilizing multi-modal affective NUI interfaces for adapting user interfaces according to user's mental and physical state. According to the authors, the analysis of bio-signals can reveal a user's preference, perception and attitude towards certain types of interfaces: colors, fonts, text size, size of the icons, the position of the elements and other characteristics of interest. Extrapolating the proposed idea, one could think of universal HCI framework that can become a widely-adopted benchmark for user acceptance of certain HMI modality. For example, innovations in MAR in the area of UI and UX can be assessed on the basis of users' anxiety, enjoyment or satisfaction.

## 5 CHALLENGES AND OPEN PROBLEMS

NUI modalities are facing several challenges. For example, BCI suffers from the low strength of the acquired biological signals, because they are recorded indirectly; localization of the source of the certain cortical activity is a challenge in EEG due to the smearing of the signal. High quality signal amplification can address this problem, improving the precision of the analysis. The wide adoption of invasive interfaces is very unlikely, but there exist commercial-driven research in the direction of embedding wires under the skin surgically for precise high-speed BCIs [119]. However, user safety and social acceptance are the key concerns in that case. Additionally, the performance of such BCI will deteriorate over time as biological tissues will inevitable die off.

Shifting and non-constant nature of biological signals, recording errors, interference and inaccurate signal classification contributes to overall high error rates. Given that, ITRs of most BCI systems are very low compared to classical I/O methods, thus minimizing errors on acquisition and pre-processing stages becomes crucial. Furthermore, the so-called "Midas touch" problem also affects most of NUI modalities. Introduced in [83], it is described as unwanted input by user when performing a natural non-HCI related activity. In eye-tracking systems, user might move his eyes away from computer screen, yet the tracking system will recognize some of the eye gestures for input. The deliberate interaction of eye movements should be distinguishable by the computer system. Similarly, intrinsic processes of thought and motor control will affect the performance of mobile BCI in unconstrained scenarios. Several solutions have been proposed for gaze tracking systems [54, 153] since this problem can be addressed by HCI applications and protocols.

In addition to technical problems of NUI, several ethical concerns regarding NUI were raised as well [86]. EEG, for example, records signals of the whole brain including cognitive activities, i.e. thinking. Thus some people might not approve their EEG data being accessed by anybody else. Similarly, EMG data can be used to identify subjects, inflating the privacy issues of NUI. Nevertheless, there are areas where NUI can hardly be replaced. Rehabilitation methods, prosthetic devices and assistive technologies heavily rely on EEG and EMG recordings especially in a case of locked in patients. Consecutively, NUIs is the only option for disabled to interact with computing systems, including mobile devices. Effective and natural control of humanoid robots and exoskeletons can be implemented by utilizing various NUI modalities [124, 155].

## 5.1 Future Directions

Traditional BCIs are moving outside of medical and research areas offering exciting opportunities for consumers. For example, combining NUIs and AR/MR/VR can make virtual worlds even more immersive by alleviating the need for physical controllers [21, 144].

***Redefining Interfaces***. Currently existing NUIs with low ITRs, such as BCIs, can be used to provide an input of several control commands in mobile HCI scenarios: 2 labels in accept/decline call; replying instant messages or emails with on of the several predefined templates; item selection in menus; feed scrolling with navigation UI elements (e.g. SSVEPs markers in the edges of the screen). In recent years, there is a trend towards lean and simplified graphical interfaces on the mobile headsets and wearables [92]. Limited bandwidth of input devices becomes a constraint to the design of a GUI. Complicated interfaces with multiple densely placed objects are not favourable for mobile scenarios. Thus, a lot of research opportunities exist for designing NUI-driven user interfaces. On the other hand, faster NUIs, like EMG, see Figure 14, can be used as alternative text input methods, swift navigation within GUI or virtual spaces. Advancements in SSI, once this technology becomes more convenient, accurate and socially accepted by the community, may lead to the formation of synthetic telepathy networks, where intended speeches (even within limited vocabulary) can be effectively obtained and securely transmitted to a recipient or broadcast to a group of audiences [114].

***Optimizing Interaction***. In 2018, the smartphone users on average spent 3 hours and 15 minutes per day on their smartphones, while the top 20% of smartphone users have been recorded the daily screen time more than 4.5 hours [109]. Among this time, 2 hours and 22 minutes on average are spent on socializing online through the six most popular platforms including Facebook, Twitter, and Instagram [40]. These social media apps on smartphones have organised the social contents on the reverse chronological timelines and the swipe gesture enables the smartphone users to navigate the contents and to scroll over news feeds horizontally on the timelines [147]. Similarly, the size-constrained Google Glass [78], which is the first-ever commercial AR head-worn computer debuted in 2013, can only accommodate a tiny touchpad on the spectacle frame of the smart glasses, and hence the operating system displays the information using *Timeline*, in which the users swipe over the pixel cards horizontally and select the targets. Thus, one of the future directions is to investigate the timeline-based interaction supported by low-ITR NUIs, and BCIs in particular. While the traditional interface design of BCI system mainly relies on the point-and-click interaction paradigm for target acquisition [125], novel interfaces can employ scroll-driven timelines. It is important to note that the scroll-driven timeline has limited capabilities to handle very complicated and voluminous information. The display of such an information can be further optimized by the context-aware architectures [92].

| Study | Modality | Application | Applied classifier | Accuracy |
|---|---|---|---|---|
| [86] | SSI | HCI | 1D CNN: 3 CL, 1 FCL, softmax | 92% |
| [161] | EMG | Prosthesis | 2D CNN : 1 CL, 2 FCL, softmax | 65% |
| [48] | EMG | Prosthesis | 2D CNN: 5 CL, 1 FCL, softmax | 66% |
| [154] | SSI | HCI | DNN: 4 FCL + HMM | N/A |
| [121] | BCI, SSVEPs | Speller | 1D CNN: 4 CL, 1 FCL, softmax | 99% |
| [166] | BCI, MI | Robot control | LSTM classifier | 93% |
| [50] | EMG | HCI | CNN: 2 CL, 1 FCL, 3 LSTM, softmax | 97% |
| [165] | BCI , MI | HCI, Speller | CNN: 2 CL, 2 FCL, softwmax + RNN: 4 FCL, 2 LSTM cells, softmax + 3 FCL | 95.53% |
| [168] | EOG | HCI, Drowsiness | CNN: 2 CL + linear regression | N/A |

Table 2. Deep learning classifiers for bio-signal processing. CNN - Convolutional Neural Network; CL - convolutional layer; FCL - Fully Connected layer; RNN - Recurrent Neural Network;

***Fashionable ExG***. Trends show that ExG sensors can be embedded into smart wearables, and several commercial products already exist employing ECG sensor mesh on the sport T-shirts [120]. Smart bands, watches, other wearable devices and even jewelry may be equipped with ExG sensors in the future. ExG electrodes can be seamlessly introduced into sports or driving helmets and glasses, providing distraction free inputs and state-monitoring (e.g. drowsiness); or incorporated into winter headdress for places with cold climate when touch-enabled input is ineffective due to low temperatures and voice interfaces are obstructed by garments [85].

***ExG on Electronic Textile***. The research community proposes Electronic textile (E-textile) and the use-case scenario is primarily smart garment. Among the most recent works, the embedded sensors in E-textiles such as strain [65] or kinetic [72] sensing and conductive stretchable fabrics [103] support the interaction between human users and mobile computers through parsing the body gestures into stitch geometries or joint movement. E-textile can also serve as a large-area touch sensible areas on smart garments [156], in order to sustain the decade-old touchscreen interaction on AR/VR head-worn computers. More importantly, ExG sensors can be designed as flexible textile electrodes [38] embedded in the spacious smart garment and consequently the awful electrodes are concealed. Through considering the social acceptance to the form-factor of electrodes, the ExG-driven interfaces can dive into the commercial markets. The form-factor and appearance characteristics of these flexible electrodes can be further evaluated as the consumer perception of the flexible electrodes can influence their popularity, for instance, a crowd-sourced survey assessing the concealed and flexible electrodes on a cap (EEG) or a glove (EMG), as well as investigating the appropriate form-factors to enhance to user awareness of the existence of ExG-enabled garments.

***Sensors Towards High Mobility***. Current issues and challenges of NUI hardware can be addressed by creating sensors that are smaller, more universal, more accurate, and more resistant to noise. Thus the development and adoption of NUI-enabled solutions might be pushed further. Moreover, progress of mobile batteries can make it possible for sensors to work autonomously for extended periods of time, this can provide substantial increase in mobility for patients that are tied to their medical equipment. In addition, progress in the technologies of wireless communications can lead to manufacturing of less energy demanding sensors that are more flexible, and require less attention from user to reconfigure connections and seamlessly add new sensors.

***Multi-sensory Signal Aggregation***. Several other conceptual solutions can be introduced to improve of NUIs. From a biological perspective EEG, EMG and EOG signals are different, but can there be an universal software platform to effectively analyze these signals? That will help to smoothly integrate different modalities into more effective hybrid interfaces. If, for example, electrodes location can be detected by properties of the acquired signal, users will be spared from burden of tedious configuration. Will the ongoing process of hardware miniaturization and further advances in machine learning techniques lead to the creation of more accurate and more precise sensors? Such a development would push NUI-enabled solutions further towards wide acceptance and usage.

***Deep Learning Approach to Bio-signal Processing***. The area of bio-signal processing is wast and historically mature. The history of BCI, for example, is almost 50 years, with first studies conducted in early 1970s [133]. Review of state-of-the-art algorithms used in bio-signal processing, such as Fourier Analysis, Power Spectrum Density (PSD) feature extraction alongside with many others, can be found in [42]. In recent years machine learning has revolutionised multiple areas including computer vision, natural language processing and ubiquitous computing. Advanced deep neural networks found their ways into self-driving cars, autonomous flying drones, a variety of IoT devices and many more. Deep learning approach has been adopted recently in the area of BCI and other bio-signal processing as well; an overview of deep learning techniques applied for healthcare and physiological signal processing is presented in [68]. Works, discussed in this survey, that employ deep learning classifiers are presented in Table 2.

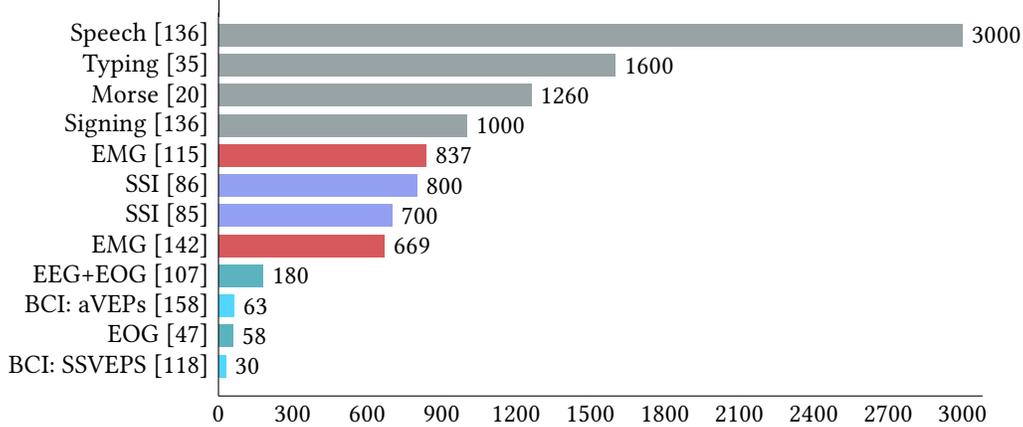

Fig. 14. Information transfer rate (ITR), in bits per minute, of communication modalities; Showing decreasing ITR from traditional interfaces (Speech and Typing) to the NUIs

## 6 CONCLUSION

In this survey we presented an overview of emerging NUI modalities. Starting from the biological foundations, we have explained the genesis of biological signals and the phenomena behind certain NUI paradigms: neural activity in the brain, retino-corneal potentials in the eye, myoelectric currents in muscles and speech production. Following, several NUI-related bio-signal acquisition techniques are mentioned as well as hardware for such an acquisition. We outlined the potential of mobile usage of imaging devices and the potential of combining these devices with MAR equipment. Next, discussion moved to the short overview of deep learning classifiers employed for bio-signal processing and existing software packages for organizing and running experiments in the area of NUIs. Finally, we have listed applications of NUIs: from assistive technologies to control of smart home environment and robotics. Reflecting back to the discussed metrics of ITR, we show a comparative summary of ITRs of discussed NUIs and other communication modalities, such as typing, in Figure 14. We believe such an overview will be helpful to those who are interested to start an applied research in the area of NUI. The survey might help choosing from available portable signal acquisition hardware; software that will help pipeline the flow of data within experiments; classification approaches and others.

**REFERENCES**

[1] 13e200 myobock electrode | professionals ottobock au. https://professionals.ottobock.com.au/Products/Prosthetics/Prosthetics-Upper-Limb/Adult-Terminal-Devices/13E200-MyoBock-electrode/p/13E200. (Accessed on 07/08/2019).
[2] Artinis medical systems | fnirs devices | nirs devices-brite. https://www.artinis.com/brite. (Accessed on 06/24/2019).
[3] Artinis medical systems | fnirs devices | nirs devices-octamon. https://www.artinis.com/octamon/. (Accessed on 06/24/2019).
[4] Artinis medical systems | fnirs devices | nirs devices-portalite. https://www.artinis.com/portalite. (Accessed on 06/24/2019).
[5] Auris. https://www.cognionics.net/auris. (Accessed on 06/22/2019).
[6] B-alert x24 - advanced brain monitoring. https://www.advancedbrainmonitoring.com/xseries/x24/. (Accessed on 06/22/2019).
[7] Br32s. wireless eeg system for brain-computer-interface. http://www.physio-tech.co.jp/pdf/br/br32s.pdf. (Accessed on 06/22/2019).
[8] Br8 plus. http://www.physio-tech.co.jp/products/br/br8_plus.html. (Accessed on 06/22/2019).
[9] ceegrid. http://ceegrid.com/home/. (Accessed on 06/22/2019).
[10] Dev kit eeg headband with dry electrodes. https://www.cognionics.net/copy-of-dev-kit. (Accessed on 06/22/2019).
[11] Dsi 24 - wearable sensing. https://wearablesensing.com/products/dsi-24/. (Accessed on 06/22/2019).
[12] Dsi 7 - wearable sensing. https://wearablesensing.com/products/dsi-7-wireless-dry-eeg-headset/. (Accessed on 06/22/2019).
[13] Dsi-vr300 | neurospec ag research neurosciences. https://www.neurospec.com/Products/Details/1077/dsi-vr300. (Accessed on 06/22/2019).
[14] Dting one - dting. http://www.dtingsmart.com/dting-one. (Accessed on 06/24/2019).
[15] Emotiv epoc+ 14 channel mobile eeg - emotiv. https://www.emotiv.com/product/emotiv-epoc-14-channel-mobile-eeg/. (Accessed on 06/22/2019).
[16] Emotiv insight 5 channel mobile eeg - emotiv. https://www.emotiv.com/product/emotiv-insight-5-channel-mobile-eeg/. (Accessed on 06/22/2019).
[17] Epoc flex gel sensor kit - emotiv. https://www.emotiv.com/product/epoc-flex-gel-sensor-kit/. (Accessed on 06/22/2019).
[18] Epoc flex saline sensor kit - emotiv. https://www.emotiv.com/product/epoc-flex-saline-sensor-kit/. (Accessed on 06/22/2019).
[19] g.nautilus - g.tec's wireless eeg system with active electrodes. http://www.gtec.at/Products/Hardware-and-Accessories/g.Nautilus-Specs-Features. (Accessed on 06/22/2019).
[20] List of interface bit rates - wikipedia. https://en.wikipedia.org/wiki/List_of_interface_bit_rates. (Accessed on 07/18/2019).
[21] Looxidvr. make your research limitless. https://looxidlabs.com/looxidvr/. (Accessed on 06/22/2019).
[22] Mindo-64 coral - r&d products - products - mindo | wearable & wireless eeg device. http://www.mindo.com.tw/en/goods.php?act=view&no=18. (Accessed on 06/22/2019).
[23] Mobile-128 wireless high density eeg. https://www.cognionics.net/mobile-128. (Accessed on 06/22/2019).
[24] Muscleban be spec sheet.pdf. https://www.biosignalsplux.com/datasheets/MuscleBAN_BE_Spec_Sheet.pdf. (Accessed on 06/24/2019).
[25] Muse 2: Brain sensing headband - technology enhanced meditation. https://choosemuse.com/muse-2/. (Accessed on 06/22/2019).
[26] Muse the brain sensing headband - technology enhanced meditation. https://choosemuse.com/muse/. (Accessed on 06/22/2019).
[27] Openbci - open source biosensing tools (eeg, emg, ekg, and more). https://openbci.com/. (Accessed on 06/22/2019).
[28] Oymotion. https://oymotion.github.io/. (Accessed on 06/24/2019).
[29] Product - zeto inc. https://zeto-inc.com/device/. (Accessed on 06/28/2019).
[30] Products / starstim / starstim nirs - neuroelectrics. https://www.neuroelectrics.com/products/starstim/starstim-nirs/. (Accessed on 06/24/2019).
[31] Quick-30 wireless dry eeg headset with dry electrodes. https://www.cognionics.net/quick-30. (Accessed on 06/22/2019).

[32] Research tools. https://store.neurosky.com/products/mindset-research-tools. (Accessed on 06/22/2019).

[33] Smarting device - high-end fully mobile eeg devices - mbraintrain. https://mbraintrain.com/smarting/. (Accessed on 06/22/2019).

[34] Tobii pro glasses 2. product description. https://www.tobiipro.com/siteassets/tobii-pro/product-descriptions/tobii-pro-glasses-2-product-description.pdf. (Accessed on 07/05/2019).

[35] Typing - wikipedia. https://en.wikipedia.org/wiki/Typing. (Accessed on 07/18/2019).

[36] Welcome to myo support. https://support.getmyo.com/hc/en-us. (Accessed on 06/24/2019).

[37] AB, T. Tobii: The world leader in eye tracking. https://www.tobii.com/, 2019. (Accessed on 10/17/2019).

[38] Acar, G., Ozturk, O., Golparvar, A. J., Elboshra, T. A., BÃűhringer, K., and Yapici, M. K. Wearable and flexible textile electrodes for biopotential signal monitoring: A review. *Electronics 8*, 5 (2019).

[39] Adimulam, M. K., and Srinivas, M. Modeling of exg (ecg, emg and eeg) non-idealities using matlab. In *2016 9th International Congress on Image and Signal Processing, BioMedical Engineering and Informatics (CISP-BMEI)* (2016), IEEE, pp. 1584–1589.

[40] Aleksandar, S. How Much Time Do People Spend on Social Media in 2019? https://techjury.net/blog/time-spent-on-social-media/, 2019. (Accessed on 10/21/2019).

[41] Alexa Internet, I. Alexa. https://www.alexa.com/, 2019. (Accessed on 10/17/2019).

[42] Alnemari, M. *Integration of a Low Cost EEG Headset with The Internet of Thing Framework*. PhD thesis, UC Irvine, 2017.

[43] Angrisani, L., Arpaia, P., Esposito, A., and Moccaldi, N. A wearable brain-computer interface instrument for augmented reality-based inspection in industry 4.0. *IEEE Transactions on Instrumentation and Measurement* (2019).

[44] Apple, I. AR Kit 3 - Apple Developers. https://developer.apple.com/augmented-reality/, 2019. (Accessed on 10/17/2019).

[45] Apple, I. Siri. https://www.apple.com/siri/, 2019. (Accessed on 10/17/2019).

[46] Arduino. Arduino. https://www.arduino.cc/, 2019. (Accessed on 10/17/2019).

[47] Atique, M. M. U., Rakib, S. H., and Siddique-e Rabbani, K. An electrooculogram based control system. In *2016 5th International Conference on Informatics, Electronics and Vision (ICIEV)* (2016), IEEE, pp. 809–812.

[48] Atzori, M., Cognolato, M., and Muller, H. Deep learning with convolutional neural networks applied to electromyography data: A resource for the classification of movements for prosthetic hands. *Frontiers in Neurorobotics 10* (2016), 9.

[49] Avalos-Viveros, H., Molero-Castillo, G., Benitez-Guerrro, E., and Bárcenas, E. Towards a method for biosignals analysis as support for the design of adaptive user-interfaces. *Advances in Pattern Recognition*, 9.

[50] Becker, V., Oldrati, P., Barrios, L., and Sörös, G. Touchsense: classifying finger touches and measuring their force with an electromyography armband. In *Proceedings of the 2018 ACM International Symposium on Wearable Computers* (2018), ACM, pp. 1–8.

[51] Bermejo, C., and Hui, P. A survey on haptic technologies for mobile augmented reality. *arXiv preprint arXiv:1709.00698* (2017).

[52] BITalino. Project BITalino. https://bitalino.com/en/, 2019. (Accessed on 10/17/2019).

[53] BTS. BTS Bioengineering. https://www.btsbioengineering.com/?page_id=12786, 2019. (Accessed on 10/17/2019).

[54] Bulling, A., Roggen, D., and Tröster, G. *Wearable EOG goggles: eye-based interaction in everyday environments*. ACM, 2009.

[55] Burnet, D. H., and Turner, M. D. Expanding eeg research into the clinic and classroom with consumer eeg systems.

[56] Chatzopoulos, D., Bermejo, C., Huang, Z., Butabayeva, A., Zheng, R., Golkarifard, M., and Hui, P. Hyperion: A wearable augmented reality system for text extraction and manipulation in the air. In *Proceedings of the 8th ACM on Multimedia Systems Conference* (New York, NY, USA, 2017), MMSys'17, ACM, pp. 284–295.

[57] Chatzopoulos, D., Bermejo, C., Huang, Z., and Hui, P. Mobile augmented reality survey: From where we are to where we go. *IEEE Access 5* (2017), 6917–6950.

[58] Chatzopoulos, D., and Hui, P. Readme: A real-time recommendation system for mobile augmented reality ecosystems. In *Proceedings of the 24th ACM International Conference on Multimedia* (2016), MM '16, pp. 312–316.

[59] Chaudhuri, A., Dasgupta, A., Chakrborty, S., and Routray, A. A low-cost, wearable, portable eog recording system. In *2016 International Conference on Systems in Medicine and Biology (ICSMB)* (2016), IEEE, pp. 102–105.

[60] Clement, J. Mobile internet usage worldwide - statistics & facts | statista. https://www.statista.com/topics/779/mobile-internet/. (Accessed on 07/06/2019).

[61] Corporation, H. Accessories: Controller. https://www.vive.com/hk/accessory/controller/, 2019. (Accessed on 10/17/2019).

[62] Deng, Y., Heaton, J. T., and Meltzner, G. S. Towards a practical silent speech recognition system. In *Fifteenth Annual Conference of the International Speech Communication Association* (2014).

[63] Dhuliawala, M., Lee, J., Shimizu, J., Bulling, A., Kunze, K., Starner, T., and Woo, W. Smooth eye movement interaction using eog glasses. In *Proceedings of the 18th ACM International Conference on Multimodal Interaction*

(2016), ACM, pp. 307–311.

[64] Diya, S. Z., Prorna, R. A., Rahman, I. I., Islam, A. B., and Islam, M. N. Applying brain-computer interface technology for evaluation of user experience in playing games. In *2019 International Conference on Electrical, Computer and Communication Engineering (ECCE)* (Feb 2019), pp. 1–6.

[65] Dupler, E., and Dunne, L. E. Effects of the textile-sensor interface on stitched strain sensor performance. In *Proceedings of the 23rd International Symposium on Wearable Computers* (2019), ISWC '19, pp. 45–53.

[66] Epic Games, I. Augmented Reality Development. https://docs.unrealengine.com/en-US/Platforms/AR/index.html, 2019. (Accessed on 10/17/2019).

[67] Fatourechi, M., Bashashati, A., Ward, R. K., and Birch, G. E. Emg and eog artifacts in brain computer interface systems: A survey. *Clinical neurophysiology 118*, 3 (2007), 480–494.

[68] Faust, O., Hagiwara, Y., Hong, T. J., Lih, O. S., and Acharya, U. R. Deep learning for healthcare applications based on physiological signals: A review. *Computer methods and programs in biomedicine 161* (2018), 1–13.

[69] Freitas, J., Ferreira, A., Figueiredo, M., Teixeira, A., and Dias, M. S. Enhancing multimodal silent speech interfaces with feature selection. In *Fifteenth Annual Conference of the International Speech Communication Association* (2014).

[70] Google, I. AR Core - Google Developers. https://developers.google.com/ar/, 2019. (Accessed on 10/17/2019).

[71] Haque, F., Nancel, M., and Vogel, D. Myopoint: Pointing and clicking using forearm mounted electromyography and inertial motion sensors. In *Proceedings of the 33rd Annual ACM Conference on Human Factors in Computing Systems* (2015), ACM, pp. 3653–3656.

[72] Hayashi, T., Ohkubo, M., Sakurai, S., Hirota, K., and Nojima, T. Towards making kinetic garments based on conductive fabric and smart hair. In *Proceedings of the 23rd International Symposium on Wearable Computers* (New York, NY, USA, 2019), ISWC '19, ACM, pp. 89–90.

[73] Herculano-Houzel, S. The human brain in numbers: a linearly scaled-up primate brain. *Frontiers in human neuroscience 3* (2009), 31.

[74] Holst, A. Number of mobile devices worldwide 2019-2023 | statistic. https://www.statista.com/statistics/245501/multiple-mobile-device-ownership-worldwide/. (Accessed on 07/06/2019).

[75] Hong, K.-S., and Khan, M. J. Hybrid brain–computer interface techniques for improved classification accuracy and increased number of commands: A review. *Frontiers in Neurorobotics 11* (2017), 35.

[76] Hyunjin Yoon, Sang-Wook Park, Yong-Kwi Lee, and Jong-Hyun Jang. Emotion recognition of serious game players using a simple brain computer interface. In *2013 International Conference on ICT Convergence (ICTC)* (Oct 2013), pp. 783–786.

[77] Inc, A. Advancer Technologies. http://www.advancertechnologies.com/, 2019. (Accessed on 10/17/2019).

[78] Inc., G. Glass Project: GLASS ENTERPRISE EDITION. https://www.google.com/glass/start/, 2019. (Accessed on 10/21/2019).

[79] Incorporated, D. Trigno Avanti Platform. https://www.delsys.com/trigno/, 2019. (Accessed on 10/17/2019).

[80] INT., I. Eye-tracking glasses: Eye-tracking with EOG. https://www.imec-int.com/en/eog, 2019. (Accessed on 10/17/2019).

[81] Ishimaru, S., Kunze, K., Uema, Y., Kise, K., Inami, M., and Tanaka, K. Smarter eyewear: using commercial eog glasses for activity recognition. In *Proceedings of the 2014 ACM International Joint Conference on Pervasive and Ubiquitous Computing: Adjunct Publication* (2014), ACM, pp. 239–242.

[82] Jackson, A. F., and Bolger, D. J. The neurophysiological bases of eeg and eeg measurement: A review for the rest of us. *Psychophysiology 51*, 11 (2014), 1061–1071.

[83] Jacob, R. J. Eye tracking in advanced interface design. *Virtual environments and advanced interface design* (1995), 258–288.

[84] Jain, J., Lund, A., and Wixon, D. The future of natural user interfaces. In *CHI'11 Extended Abstracts on Human Factors in Computing Systems* (2011), ACM, pp. 211–214.

[85] Jorgensen, C., and Dusan, S. Speech interfaces based upon surface electromyography. *Speech Communication 52*, 4 (2010), 354–366.

[86] Kapur, A., Kapur, S., and Maes, P. Alterego: A personalized wearable silent speech interface. In *23rd International Conference on Intelligent User Interfaces* (2018), ACM, pp. 43–53.

[87] Kassner, M., Patera, W., and Bulling, A. Pupil: an open source platform for pervasive eye tracking and mobile gaze-based interaction. In *Proceedings of the 2014 ACM international joint conference on pervasive and ubiquitous computing: Adjunct publication* (2014), ACM, pp. 1151–1160.

[88] Kim, Y., Doh, N. L., Youm, Y., and Chung, W. K. Robust discrimination method of the electrooculogram signals for human-computer interaction controlling mobile robot. *Intelligent Automation & Soft Computing 13*, 3 (2007), 319–336.

[89] Klemmer, S. R., Hartmann, B., and Takayama, L. How bodies matter: Five themes for interaction design. In *Proceedings of the 6th Conference on Designing Interactive Systems* (2006), DIS '06, pp. 140–149.

[90] Kos'myna, N., Tarpin-Bernard, F., and Rivet, B. Bidirectional feedback in motor imagery bcis: Learn to control a drone within 5 minutes. In *CHI '14 Extended Abstracts on Human Factors in Computing Systems* (New York, NY, USA, 2014), CHI EA '14, ACM, pp. 479–482.

[91] Lab, T. M. P. V. Eog. https://www.medicine.mcgill.ca/physio/vlab/Other_exps/EOG/eogintro_n.htm. (Accessed on 07/17/2019).

[92] Lam, K. Y., Lee, L. H., Braud, T., and Hui, P. M2a: A framework for visualizing information from mobile web to mobile augmented reality. In *2019 IEEE International Conference on Pervasive Computing and Communications (PerCom)* (March 2019), pp. 1–10.

[93] Lee, H. J., Kim, H. S., and Park, K. S. A study on the reproducibility of biometric authentication based on electroencephalogram (eeg). In *2013 6th international IEEE/EMBS Conference on Neural Engineering (NER)* (2013), IEEE, pp. 13–16.

[94] Lee, J., Cha, K., Kim, H., Choi, J., Kim, C., and Lee, S. Hybrid mi-sssep paradigm for classifying left and right movement toward bci for exoskeleton control. In *2019 7th International Winter Conference on Brain-Computer Interface (BCI)* (2019), IEEE, pp. 1–3.

[95] LEE, L.-H. Embodied Interaction on Constrained Interfaces. HKUST Online Thesis, 2019. (Accessed on 10/17/2019).

[96] Lee, L. H., Braud, T., Bijarbooneh, F. H., and Hui, P. Tipoint: Detecting fingertip for mid-air interaction on computational resource constrained smartglasses. In *Proceedings of the 23rd International Symposium on Wearable Computers* (New York, NY, USA, 2019), ISWC '19, ACM, pp. 118–122.

[97] Lee, L.-H., and Hui, P. Interaction methods for smart glasses: A survey. *IEEE Access 6* (2018), 28712–28732.

[98] Lee, L. H., Lam, K. Y., Li, T., Braud, T., Su, X., and Hui, P. Quadmetric optimized thumb-to-finger interaction for force assisted one-handed text entry on mobile headsets. *Proc. ACM Interact. Mob. Wearable Ubiquitous Technol. 3*, 3 (Sept. 2019), 94:1–94:27.

[99] Lee, L. H., Yung Lam, K., Yau, Y. P., Braud, T., and Hui, P. Hibey: Hide the keyboard in augmented reality. In *2019 IEEE International Conference on Pervasive Computing and Communications (PerCom)* (March 2019), pp. 1–10.

[100] Lin, C.-T., Ko, L.-W., Chang, M.-H., Duann, J.-R., Chen, J.-Y., Su, T.-P., and Jung, T.-P. Review of wireless and wearable electroencephalogram systems and brain-computer interfaces–a mini-review. *Gerontology 56*, 1 (2010), 112–119.

[101] Lin, Z., Zhang, C., Wu, W., and Gao, X. Frequency recognition based on canonical correlation analysis for ssvep-based bcis. *IEEE transactions on biomedical engineering 53*, 12 (2006), 2610–2614.

[102] Ling, K., Dai, H., Liu, Y., and Liu, A. X. Ultragesture: Fine-grained gesture sensing and recognition. In *2018 15th Annual IEEE International Conference on Sensing, Communication, and Networking (SECON)* (2018), IEEE, pp. 1–9.

[103] Liu, R., Shao, Q., Wang, S., Ru, C., Balkcom, D., and Zhou, X. Reconstructing human joint motion with computational fabrics. *Proc. ACM Interact. Mob. Wearable Ubiquitous Technol. 3*, 1 (Mar. 2019), 19:1–19:26.

[104] Liu, S. Consumer ar app market revenue worldwide 2016-2022 statistic. https://www.statista.com/statistics/608990/mobile-ar-applications-installed-base-worldwide-by-type/. (Accessed on 07/06/2019).

[105] Liu, S. Global augmented/virtual reality market size 2016-2023 statistic. https://www.statista.com/statistics/591181/global-augmented-virtual-reality-market-size/. (Accessed on 07/06/2019).

[106] M. Abuhasira, S. V., and Geva, A. B. s9554-fast-training-of-deep-neural-networks-using-brain-generated-labels.pdf. https://developer.download.nvidia.com/video/gputechconf/gtc/2019/presentation/s9554-fast-training-of-deep-neural-networks-using-brain-generated-labels.pdf. (Accessed on 07/03/2019).

[107] Ma, J., Zhang, Y., Cichocki, A., and Matsuno, F. A novel eog/eeg hybrid human–machine interface adopting eye movements and erps: Application to robot control. *IEEE Transactions on Biomedical Engineering 62*, 3 (2014), 876–889.

[108] Ma, Z., Li, B. C., and Yan, Z. Wearable driver drowsiness detection using electrooculography signal. In *2016 IEEE Topical Conference on Wireless Sensors and Sensor Networks (WiSNet)* (2016), IEEE, pp. 41–43.

[109] MacKay, J. Screen time stats 2019: HereâĂŹs how much you use your phone during the workday. https://blog.rescuetime.com/screen-time-stats-2018/, 2019. (Accessed on 10/21/2019).

[110] Meltzner, G. S., Heaton, J. T., Deng, Y., De Luca, G., Roy, S. H., and Kline, J. C. Silent speech recognition as an alternative communication device for persons with laryngectomy. *IEEE/ACM transactions on audio, speech, and language processing 25*, 12 (2017), 2386–2398.

[111] MEME, J. JINS MEME ES: EYE SENSING. https://jins-meme.com/en/products/es/, 2019. (Accessed on 10/17/2019).

[112] Mercier-Ganady, J., Marchal, M., and Lécuyer, A. B-c-invisibility power: Introducing optical camouflage based on mental activity in augmented reality. In *Proceedings of the 6th Augmented Human International Conference* (2015), pp. 97–100.

[113] Microsoft. Kinect for Windows. https://developer.microsoft.com/en-us/windows/kinect, 2019. (Accessed on 10/17/2019).

[114] Mohanchandra, K., Saha, S., and Lingaraju, G. Eeg based brain computer interface for speech communication: principles and applications. In *Brain-Computer Interfaces*. Springer, 2015, pp. 273–293.

[115] Moin, A., Zhou, A., Benatti, S., Rahimi, A., Alexandrov, G., Menon, A., Tamakloe, S., Ting, J., Yamamoto, N., Khan, Y., Burghardt, F., Arias, A. C., Benini, L., and Rabaey, J. M. Adaptive emg-based hand gesture recognition using hyperdimensional computing. *CoRR abs/1901.00234* (2019).

[116] Motion Lab Systems, I. Multi-Channel Electromyography Systems. https://www.motion-labs.com/, 2019. (Accessed on 10/17/2019).

[117] NA. Low cost open source eeg device completely assembled USB interface. https://www.olimex.com/Products/EEG/OpenEEG/EEG-SMT/open-source-hardware, 2019. (Accessed on 10/17/2019).

[118] Nakanishi, M., Wang, Y., Wang, Y.-T., Mitsukura, Y., and Jung, T.-P. A high-speed brain speller using steady-state visual evoked potentials. *International journal of neural systems 24*, 06 (2014), 1450019.

[119] Neuralink. Neuralink Launch Event. https://www.youtube.com/watch?v=r-vbh3t7WVI, 2019. (Accessed on 10/17/2019).

[120] Neuralink. ZEPTH: Intelligent Fitness and Sport Shirts. https://www.xiaomiyoupin.com/detail?gid=108157, 2019. (Accessed on 10/17/2019).

[121] Nguyen, T.-H., and Chung, W.-Y. A single-channel ssvep-based bci speller using deep learning. *IEEE Access 7* (2018), 1752–1763.

[122] Nishifuji, S., Nakamura, H., and Matsubara, A. Brain computer interface using modulation of auditory steady-state response with help of stochastic resonance. In *2018 40th Annual International Conference of the IEEE Engineering in Medicine and Biology Society (EMBC)* (2018), IEEE, pp. 2028–2031.

[123] Norton, J. J., Mullins, J., Alitz, B. E., and Bretl, T. The performance of 9–11-year-old children using an ssvep-based bci for target selection. *Journal of neural engineering 15*, 5 (2018), 056012.

[124] Nourmohammadi, A., Jafari, M., and Zander, T. O. A survey on unmanned aerial vehicle remote control using brain–computer interface. *IEEE Transactions on Human-Machine Systems 48*, 4 (2018), 337–348.

[125] Obbink, M., Gürkök, H., Plass-Oude Bos, D., Hakvoort, G., Poel, M., and Nijholt, A. Social interaction in a cooperative brain-computer interface game. In *Intelligent Technologies for Interactive Entertainment* (Berlin, Heidelberg, 2012), A. Camurri and C. Costa, Eds., Springer Berlin Heidelberg, pp. 183–192.

[126] OpenEEG. OpenEEG. http://openeeg.sourceforge.net/doc/index.html, 2019. (Accessed on 10/17/2019).

[127] Oskoei, M. A., and Hu, H. Myoelectric control systems survey. *Biomedical signal processing and control 2*, 4 (2007), 275–294.

[128] Paramesura Rao, V. R., Hewawasam Puwakpitiyage, C. A., Muhammad Azizi, M. S. A., Tee, W. J., Murugesan, R. K., and Hamzah, M. D. Emotion recognition in e-commerce activities using eeg-based brain computer interface. In *2018 Fourth International Conference on Advances in Computing, Communication Automation* (Oct 2018), pp. 1–5.

[129] Potts, D., Loveys, K., Ha, H., Huang, S., Billinghurst, M., and Broadbent, E. Zeng: Ar neurofeedback for meditative mixed reality. In *Proceedings of the 2019 on Creativity and Cognition* (2019), ACM, pp. 583–590.

[130] Prophet, J., Kow, Y. M., and Hurry, M. Small trees, big data: Augmented reality model of air quality data via the chinese art of "artificial" tray planting. In *ACM SIGGRAPH 2018 Posters* (2018), pp. 16:1–16:2.

[131] Putze, F. Methods and tools for using bci with augmented and virtual reality. In *Brain Art*. Springer, 2019, pp. 433–446.

[132] Radüntz, T. Signal quality evaluation of emerging eeg devices. *Frontiers in physiology 9* (2018), 98.

[133] Ramadan, R. A., Refat, S., Elshahed, M. A., and Ali, R. A. Basics of brain computer interface. In *Brain-Computer Interfaces*. Springer, 2015, pp. 31–50.

[134] Ramchurn, R., Chamberlian, A., and Benford, S. Designing musical soundtracks for brain controlled interface (bci) systems. In *Proceedings of the Audio Mostly 2018 on Sound in Immersion and Emotion* (New York, NY, USA, 2018), AM'18, ACM, pp. 28:1–28:8.

[135] Ratti, E., Waninger, S., Berka, C., Ruffini, G., and Verma, A. Comparison of medical and consumer wireless eeg systems for use in clinical trials. *Frontiers in human neuroscience 11* (2017), 398.

[136] Reed, C. M., and Durlach, N. I. Note on information transfer rates in human communication. *Presence 7*, 5 (1998), 509–518.

[137] Rezeika, A., Benda, M., Stawicki, P., Gembler, F., Saboor, A., and Volosyak, I. Brain–computer interface spellers: A review. *Brain sciences 8*, 4 (2018), 57.

[138] S.A., P. W. B. Bio-signal PLUX. https://plux.info/, 2019. (Accessed on 10/17/2019).

[139] Schultz, T., Wand, M., Hueber, T., Krusienski, D. J., Herff, C., and Brumberg, J. S. Biosignal-based spoken communication: A survey. *IEEE/ACM Transactions on Audio, Speech, and Language Processing 25*, 12 (2017), 2257–2271.

[140] Services, A. W. Amazon Sumerian: Easily create and run browser-based 3D, augmented reality (AR), and virtual reality (VR) applications. https://aws.amazon.com/sumerian/, 2019. (Accessed on 10/17/2019).

[141] Shannon, C. E. A mathematical theory of communication. *Bell system technical journal 27*, 3 (1948), 379–423.

[142] Shatilov, K. A., Chatzopoulos, D., Hang, A. W. T., and Hui, P. Using deep learning and mobile offloading to control a 3d-printed prosthetic hand. *Proc. ACM Interact. Mob. Wearable Ubiquitous Technol. 3*, 3 (Sept. 2019), 102:1–102:19.

[143] Si-Mohammed, H., Petit, J., Jeunet, C., Argelaguet, F., Spindler, F., Evain, A., Roussel, N., Casiez, G., and Lécuyer, A. Towards bci-based interfaces for augmented reality: Feasibility, design and evaluation. *IEEE transactions on visualization and computer graphics* (2018).

[144] Si-Mohammed, H., Sanz, F. A., Casiez, G., Roussel, N., and Lécuyer, A. Brain-computer interfaces and augmented reality: A state of the art. In *Graz Brain-Computer Interface Conference, Graz Austria* (2017).

[145] Soundariya, R., and Renuga, R. Eye movement based emotion recognition using electrooculography. In *2017 Innovations in Power and Advanced Computing Technologies (i-PACT)* (2017), IEEE, pp. 1–5.

[146] srl, C. PicoEMG. https://www.cometasystems.com/products/picoemg, 2019. (Accessed on 10/17/2019).

[147] Stinson, L. Stop the Endless Scroll. Delete Social Media From Your Phone. https://www.wired.com/story/rants-and-raves-desktop-social-media/, 2019. (Accessed on 10/21/2019).

[148] Tamaki, D., Fujimori, H., and Tanaka, H. An interface using electrooculography with closed eyes. In *International Symposium on Affective Science and Engineering ISASE2019* (2019), Japan Society of Kansei Engineering, pp. 1–4.

[149] Tayeb, Z., Waniek, N., Fedjaev, J., Ghaboosi, N., Rychly, L., Widderich, C., Richter, C., Braun, J., Saveriano, M., Cheng, G., et al. Gumpy: A python toolbox suitable for hybrid brain–computer interfaces. *Journal of neural engineering 15*, 6 (2018), 065003.

[150] Technologies, F. Rift: Accessories. https://www.oculus.com/rift/accessories/, 2019. (Accessed on 10/17/2019).

[151] Technologies, U. Augmented Reality (AR). https://unity.com/unity/features/ar, 2019. (Accessed on 10/17/2019).

[152] TMSi. Small wearable amplifier system. https://www.tmsi.com/products/mobi/, 2019. (Accessed on 10/17/2019).

[153] Velichkovsky, B. B., Rumyantsev, M. A., and Morozov, M. A. New solution to the midas touch problem: Identification of visual commands via extraction of focal fixations. *Procedia Computer Science 39* (2014), 75–82.

[154] Wand, M., and Schmidhuber, J. Deep neural network frontend for continuous emg-based speech recognition. In *INTERSPEECH* (2016), pp. 3032–3036.

[155] Wang, M., Li, R., Zhang, R., Li, G., and Zhang, D. A wearable ssvep-based bci system for quadcopter control using head-mounted device. *IEEE Access 6* (2018), 26789–26798.

[156] Wang, Y., Zhou, J., Li, H., Zhang, T., Gao, M., Cheng, Z., Yu, C., Patel, S., and Shi, Y. Flextouch: Enabling large-scale interaction sensing beyond touchscreens using flexible and conductive materials. *Proc. ACM Interact. Mob. Wearable Ubiquitous Technol. 3*, 3 (Sept. 2019), 109:1–109:20.

[157] Wu, S.-L., Liao, L.-D., Lu, S.-W., Jiang, W.-L., Chen, S.-A., and Lin, C.-T. Controlling a human–computer interface system with a novel classification method that uses electrooculography signals. *IEEE transactions on Biomedical Engineering 60*, 8 (2013), 2133–2141.

[158] Xu, M., Xiao, X., Wang, Y., Qi, H., Jung, T.-P., and Ming, D. A brain–computer interface based on miniature-event-related potentials induced by very small lateral visual stimuli. *IEEE Transactions on Biomedical Engineering 65*, 5 (2018), 1166–1175.

[159] Yoshioka, M., Inoue, T., and Ozawa, J. Brain signal pattern of engrossed subjects using near infrared spectroscopy (nirs) and its application to tv commercial evaluation. In *The 2012 International Joint Conference on Neural Networks (IJCNN)* (June 2012), pp. 1–6.

[160] Yu, N., Wang, W., Liu, A. X., and Kong, L. Qgesture: Quantifying gesture distance and direction with wifi signals. *Proceedings of the ACM on Interactive, Mobile, Wearable and Ubiquitous Technologies 2*, 1 (2018), 51.

[161] Zhai, X., Jelfs, B., Chan, R. H. M., and Tin, C. Self-recalibrating surface emg pattern recognition for neuroprosthesis control based on convolutional neural network. *Frontiers in Neuroscience 11* (2017), 379.

[162] Zhang, J., Wang, B., Zhang, C., Xiao, Y., and Wang, M. Y. An eeg/emg/eog-based multimodal human-machine interface to real-time control of a soft robot hand. *Frontiers in neurorobotics 13* (2019).

[163] Zhang, W., Han, B., and Hui, P. On the networking challenges of mobile augmented reality. In *Proceedings of the Workshop on Virtual Reality and Augmented Reality Network* (2017), ACM, pp. 24–29.

[164] Zhang, W., Han, B., and Hui, P. Jaguar: Low latency mobile augmented reality with flexible tracking. In *2018 ACM Multimedia Conference on Multimedia Conference* (2018), ACM, pp. 355–363.

[165] Zhang, X., Yao, L., Sheng, Q. Z., Kanhere, S. S., Gu, T., and Zhang, D. Converting your thoughts to texts: Enabling brain typing via deep feature learning of eeg signals. In *2018 IEEE International Conference on Pervasive Computing and Communications (PerCom)* (2018), IEEE, pp. 1–10.

[166] Zhang, X., Yao, L., Zhang, S., Kanhere, S., Sheng, M., and Liu, Y. Internet of things meets brain–computer interface: A unified deep learning framework for enabling human-thing cognitive interactivity. *IEEE Internet of Things Journal 6*, 2 (2018), 2084–2092.

[167] Zhang, Y., Gu, T., Luo, C., Kostakos, V., and Seneviratne, A. Findroidhr: Smartwatch gesture input with optical heartrate monitor. *Proceedings of the ACM on Interactive, Mobile, Wearable and Ubiquitous Technologies 2*, 1 (2018), 56.

[168] Zhu, X., Zheng, W.-L., Lu, B.-L., Chen, X., Chen, S., and Wang, C. Eog-based drowsiness detection using convolutional neural networks. In *2014 International Joint Conference on Neural Networks (IJCNN)* (2014), IEEE, pp. 128–134.

# APPENDIX A - PORTABLE EEG HEADSETS

| Device: | **1.** EMOTIV EPOC+ [15] 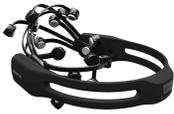 | **2.** EMOTIV Insight 5 [16] 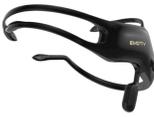 | **3.** AEPOC Flex Saline Sensor Kit [18] 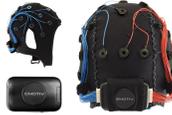 | **4.** EPOC Flex Gel Sensor Kit [17] 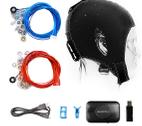 |
|---|---|---|---|---|
| Electrodes: | 14 wet | 5 semi-dry | 32 wet | 32 wet |
| Channels: | AF3, F7, F3, FC5, T7, P7, O1, O2, P8, T8, FC6, F4, F8, AF4 | AF3, AF4, T7, T8, Pz | Configurable | Configurable |
| Additional Sensors: | Magnetometer, Accelerometer | Gyroscope, Magnetometer, Accelerometer | Magnetometer, Accelerometer | Magnetometer, Accelerometer |
| Resolution: | 14 or 16 and 16 bits | 14 bits and 14 bits | 14 bits | 14 bits |
| Sampling rate: | 2048 → 256 or 128 | 128 | 1024 → 128 | 1024 → 128 |
| Protocol: | BLE | BLE | Proprietary wireless protocol @ 2.4Ghz | Proprietary wireless protocol @ 2.4Ghz |
| Battery: | 640mAh: 12h (USB receiver), 6h BLE | 480mAh: 8h (USB receiver), 4h BLE | 640mAh: 9h | 640mAh: 9h |
| Device: | **5.** Neurosky [32] 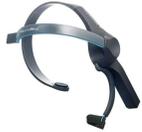 | **6.** Muse (2014) [26] 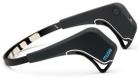 | **7.** Muse 2 (2016) [25] 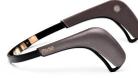 | **8.** mBrainTrain SMARTING 24 [33] 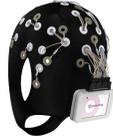 |
| Electrodes: | 1 dry | 4 dry | 4 dry | 24 dry |
| Channels: | FP1 | TP9, AF7, AF8, TP10 | TP9, AF7, AF8, TP10 | FP1, FP2, AFz, F7, F3, Fz, F4, F8, T7, C3, Cz, C4, T8, M1, CPz, M2, P7, P3, Pz, P4, P8, POz, O1, O2 |
| Additional Sensors: | - | Accelerometer | PPG, Accelerometer | Gyroscope |
| Resolution: | 12 bits | 10 or 16 bits | 10 or 16 bits | 24 bits |
| Sampling rate: | 512 | 220Hz or 500Hz | 256Hz | 250Hz or 500 Hz |
| Protocol: | BT/BLE | BT 2.1 + EDR | BT 4.0 and BLE | BT 2.1 + EDR |
| Battery: | 1000mAh (AAA Battery): 8-hours | 5h | 5h | 560 mAh: 4h |

Table 3. Portable EEG headsets. Part 1.

| Device: | 9. cEEGrid [9] 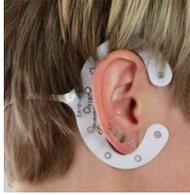 | 10. Cognionics QUICK 8/20/30 [31] 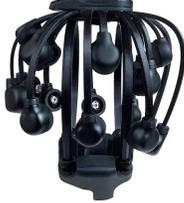 | 11. Cognionics Mobile 64/128 [23] 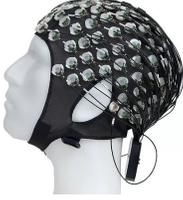 | 12. Auris In-Ear EEG [5] 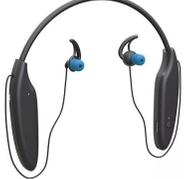 |
|---|---|---|---|---|
| Electrodes: | 18 dry | 8/20/30 dry | 64/128 wet | 2 dry (+6 via extension) |
| Channels: | Custom locations around ears | Configurable (hot placement) | Configurable | Left, Right ear and Chest ECG |
| Additional Sensors: | Gyroscope | - | Accelerometer, Gyroscope | - |
| Resolution: | 24 bits | 24 bits | 24 bits | 24 bits |
| Sampling rate: | 250 or 500Hz | 500Hz or 1000Hz or 2000Hz | 500Hz or 1000Hz | 250 or 500 or 1,000 or 2,000 Hz |
| Protocol: | BT 2.1 | BT | BT | BT |
| Battery: | - | 8h (wireless); 16h (microSD card) | 6h (wireless); 8h (microSD card) | 8h (wireless); 16h (microSD card) |

| Device: | 13. Cognionics DevKit [10] 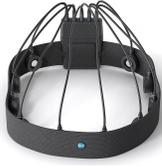 | 14. Open BCI Ultracortex "Mark IV" EEG Headset + Cyton Biosensing Board [27] 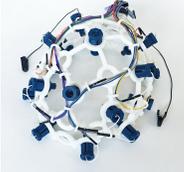 | 15. DSI-VR300 [13] 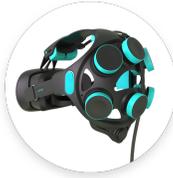 | 16. LooxidVR Headset & Mask [21] 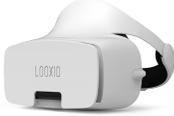 |
|---|---|---|---|---|
| Electrodes: | 8 dry | 8 dry (+8 via extension) | 7 dry | 6 dry |
| Channels: | Configurable around the headband | Configurable, chassis can be 3D printed | Fz, Pz, P3, P4, PO7, PO8, Oz | 6 electrodes over the frontal lobe, custom positions |
| Additional Sensors: | Configurable | Accelerometer | Optional embedded 3D accelerometers | Eye tracking camera (eye movement, and pupil dilation), IMU |
| Resolution: | 24 bits | 24 bits | 16 bits | 24 bits |
| Sampling rate: | 250 or 500 or 1,000 or 2,000 Hz | 250 HZ | 300 Hz (600 Hz option) | 1000 Hz |
| Protocol: | BT | BLE | BT | BT |
| Battery: | 8h (wireless); 16h (microSD card) | 4000 mAh (4 AAA batteries) | 12+ h | Requires external source (USB +5V DC) |

Table 4. Portable EEG headsets. Part 2.

| Device: | **17.** B-Alert X24/X10 EEG System [6] 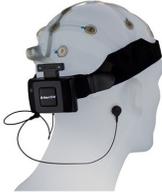 | **18.** BR32S [7] 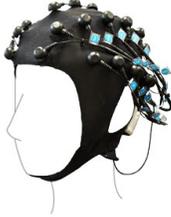 | **19.** BR8PLUS [8] 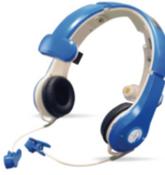 | **20.** Mindo Jellyfish/Trilobite/ Coral [22] 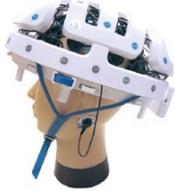 |
|---|---|---|---|---|
| Electrodes: | 9+1/20+4 | 32 | 8 | 4/32/64 |
| Channels: | Fz, F3, F4, Cz, C3, C4, POz, P3, P4 / Fz, F1, F2, F3, F4, Cz, C1, C2, C3, C4, CPz, Pz, P1, P2, P3, P4, POz, Oz, O1, O2 | 32 10-20 system | Fp1, Fp2, Fz, C3, C4, Pz, O1, O2 | AF7, Fp1, Fp2, AF8 / 32 10-20 system / 64 10-20 system |
| Additional Sensors: | 3-axis accelerometer | - | - | - |
| Resolution: | 16 bits and 12 bits | 16 bits | 16 bits | 24 bits |
| Sampling rate: | 256Hz | 250Hz | 250Hz | 500Hz |
| Protocol: | BT | BT 2.1 | BT 2.1 | BT |
| Battery: | 8+ h | 10+ h | 10+ h | - |

| Device: | **21.** DSI 24 [11] 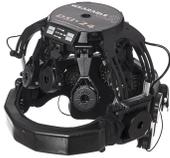 | **22.** DSI 7 (DSI flex) [12] 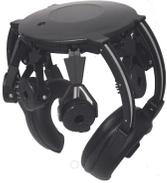 | **23.** g.Nautilus (PRO) [19] 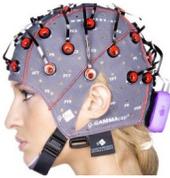 | **24.** Zeto [29] 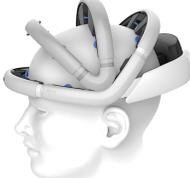 |
|---|---|---|---|---|
| Electrodes: | 21 dry | 7 dry | 8/16/32 dry active | 19 dry |
| Channels: | Fp1, Fp2, Fz, F3, F4, F7, F8, Cz, C3, C4, T7/T3, T8/T4, Pz, P3, P4, P7/T5, P8/T6, O1, O2, A1, A2 | F3, F4, C3, C4, Pz, P3, P4 (Configurable for DSI flex) | Fp1, Fp2, F3, Fz, F4, P3, Pz, P4, PO7, Oz, PO8 / 32 10-20 system / 64 10-20 system | Fp1, Fp2, F7, F3, Fz, F4, F8, T3, C3, Cz, C4, T4, T5„ P3, Pz, P4, T6, O1, O2 |
| Additional Sensors: | Optional embedded 3D accelerometer | Optional embedded 3D accelerometer | 3-axis accelerometer | None |
| Resolution: | 16 bits | 16 bits | 24 bits | 24 bits |
| Sampling rate: | 300 Hz (600 Hz option) | 300 Hz | 500 Hz | 500 Hz |
| Protocol: | BT | BT | Proprietary wireless protocol @ 2.4Ghz | 2.4 GHz WiFi, 802.11 b/g/n |
| Battery: | 8+ h | 8+ h | 10+ h | 6-7 h |

Table 5. Portable EEG headsets. Part 3.

## APPENDIX B - PORTABLE FNIRS SYSTEMS

| Device: | **1.** Brite [2] | **2.** PortaLite [4] | **3.** Starstim fNIRS (Combined with tCS and EEG) [30] | **4.** Octa-Mon/OctaMon+ [3] |
|---|---|---|---|---|
| | 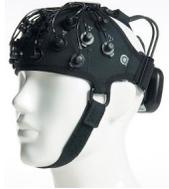 | 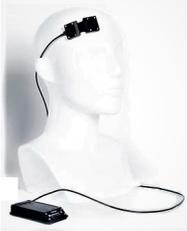 | 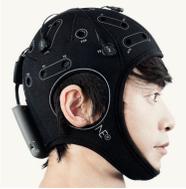 | 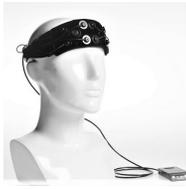 |
| **Channels:** | up to 54 | 4 | 8 or 24 | 8 |
| **Resolution:** | 16 bits | 16 bits | EEG: 24 bits | 16 bit |
| **Sampling rate:** | 50 or 100 Hz | 50 Hz | fNIRS: 50 Hz; EEG: 500 Hz | 10 Hz |
| **Protocol:** | BT | BT | BT | BT |
| **Battery:** | 3+ hours (power bank compatible) | 8 or 16 h | - | 6h / 3-11h |

Table 6. Portable fNIRS systems.

## APPENDIX C - WIRELESS EMG SENSORS

| Device: | **1.** MYO Band [36] | **2.** Gforce OyMotion Band (PRO) [28] | **3.** DTing One [14] | **4.** MuscleBANBE [24] |
|---|---|---|---|---|
| | 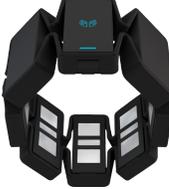 | 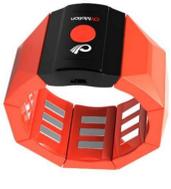 | 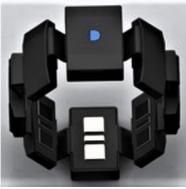 | 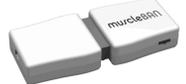 |
| **Channels:** | 8 | 8 | 8 | 1 |
| **Additional Sensors:** | 9-Axis IMU, 13 bit | 9-Axis, Accelerometer, Gyroscope, Magnetometer 50Hz, 16bit | Accelerometer, Gyroscope: 200Hz, 16bits | Accelerometer: 14bits; Magnetometer: 16bits |
| **Resolution:** | 16bits | 8bits | 12bits | 12bits |
| **Sampling rate:** | 200Hz | 1000Hz | 1000Hz | 1000Hz |
| **Protocol:** | BLE | BLE4.1 | BLE 5.0, Wi-Fi | BLE |
| **Battery:** | 2 x 260mAh | 200mAh | - | 155 mAh: 8h |

Table 7. Wireless EMG sensors.